\documentclass[final,3p,times,twocolumn]{elsarticle}
\biboptions{comma,sort&compress}
\usepackage{ecrc}
\usepackage{here}
\usepackage{graphicx}
\usepackage{epsfig}
\usepackage{epstopdf}
\usepackage{amsmath}
\def\nin{\noindent}
\def\beq{\begin{equation}}
\def\eeq{\end{equation}}
\def\bea{\begin{eqnarray}}
\def\eea{\end{eqnarray}}
\def\nnb{\nonumber}
\def\la{\langle}
\def\ra{\rangle}
\def\ga{\left(}
\def\dr{\right)}

\usepackage{graphicx}
\usepackage{here}
\def\beq{\begin{equation}}
\def\eeq{\end{equation}}
\def\bea{\begin{eqnarray}}
\def\eea{\end{eqnarray}}
\def\bq{\begin{quote}}
\def\eq{\end{quote}}

\def\nnb{\nonumber}
\def\ga{\left(}
\def\dr{\right)}

\def\nnb{\nonumber}
\def\la{\langle}
\def\ra{\rangle}
\def\nin{\noindent}
\def\ba{\begin{array}}
\def\ea{\end{array}}

\def\als{\alpha_s}

\def\gg2{ \la\alpha_s G^2 \ra}
\def\gg3{g^3f_{abc}\la G^aG^bG^c \ra}
\def\ggg4{\la\als^2G^4\ra}

\def\beq{\begin{equation}}
\def\enq{\end{equation}}
\def\beqa{\begin{eqnarray}}
\def\enqa{\end{eqnarray}}
\def\nnb{\nonumber}

\def\qq{\lag\bar{q}q\rag}

\def\lb{\label}

\newcommand{\rag}{\rangle}
\newcommand{\lag}{\langle}

\def\ln{\mbox{Log}}
\def\gg{\lag g^{2}_{s} G^2 \rag}
\def\ggg{\lag g^{3}_{s}G^3\rag}
\volume{00}
\firstpage{1}
\journalname{Nuclear and Particle Physics Proceedings }
\runauth{Albuquerque et al.}
\jnltitlelogo{Nuclear and Particle Physics Proceedings }

\begin{document}
\begin{frontmatter}

\title{ XYZ -- spectra from QCD Laplace Sum Rules at Higher Orders \tnoteref{text1}}

%\author{D. Rabetiarivony\corref{cor2}\fnref{label1}}
%\address{Institute of High-Energy Physics of Madagascar (iHEPMAD), University of Antananarivo, Madagascar}
%\ead{rd.bidds@gmail.com}
%\fntext[label1]{Ph.D student}
%\tnotetext[text1]{Talk given at  HEPMAD17 (21 - 26 September, Antananarivo - Madagascar)}
%\cortext[cor2]{Speaker}

\author{R. Albuquerque}
\address{Faculty of Technology,Rio de Janeiro State University (FAT,UERJ), Brazil}
\ead{raphael.albuquerque@uerj.br}
\author{S. Narison\corref{cor1}
}
\address{Laboratoire
Univers et Particules de Montpellier (LUPM), CNRS-IN2P3, \\
Case 070, Place Eug\`ene
Bataillon, 34095 - Montpellier, France.}
\ead{snarison@yahoo.fr}

\author{A. Rabemananjara%\corref{cor2}
%\fnref{label1}
}\ead{achrisrab@gmail.com}

\author{D. Rabetiarivony%\corref{cor2}
\fnref{label1}}\ead{rd.bidds@gmail.com}
\author{G. Randriamanatrika\fnref{label1}}\ead{artesgaetan@gmail.com}

\address{Institute of High-Energy Physics of Madagascar (iHEPMAD)\\
University of Ankatso,
Antananarivo 101, Madagascar}

\fntext[label1]{Ph.D student}
\tnotetext[text1]{Talk presented at  QCD18 (2--6 july 2018, Montpellier--FR) and HEPMAD18 (6--11 September 2018, Antananarivo--MG)}
\cortext[cor1]{ICTP-Trieste high-energy physics consultant for Madagascar}
%\cortext[cor2]{Speaker at HEPMAD17}


% \address[label1]{Institute of High Energy Physics of Madagascar (iHEP-MAD), University of Antananarivo, Madagascar }
%\ead{fanfenos@yahoo.fr}
%\address[label2]{Laboratoire
%Univers et Particules (LUPM), CNRS-IN2P3 \& Universit\'e
%de Montpellier II, 
%\\
%Case 070, Place Eug\`ene
%Bataillon, 34095 - Montpellier Cedex 05, France.}
%\cortext[cor1]{Speaker}
%\ead{snarison@yahoo.fr}

%\author[label1]{A. Rabemananjara\corref{cor2}}
%  \address[label1]{Institute of High Energy Physics of Madagascar iHEP-MAD, University of Antananarivo, Madagascar  }
%\cortext[cor1]{Talk given at  QCD 14 (29 june - 3 july, Montpellier - France).}
%\cortext[cor2]{Speaker.}
%\ead{{achris$\_$}01@yahoo.fr.}
%\author{}

%\address{}

\begin{abstract}
\noindent
We review our results in Refs.\,\cite{SU3,SNX2} for the masses and couplings of heavy-light $\bar DD(\bar BB)$-like molecules and $(\overline{Qq})(Qq)$-like four-quark  states from relativistic QCD Laplace sum rules (LSR) where next-to-next-to-leading order (N2LO) PT corrections in the chiral limit, next-to-leading order (NLO) SU3 PT corrections and non-perturbative contributions up to dimension $d=6-8$ are included.  The factorization properties of molecule and four-quark currents have been used for the estimate of the higher order PT corrections.
New integrated compact expressions of the spectral functions at leading order (LO) of perturbative QCD and up to dimensions $d\leq (6 - 8)$ non-perturbative condensates are presented .
%use QCD spectral sum rules (QSSR) and the factorization properties of molecule and four-quark currents to estimate  the masses and couplings of some molecules and four-quark  states, including next-to-next-to-leading order (N2LO) PT corrections in the chiral limit and next-to-leading order (NLO) SU3 PT corrections. 
%We use new integrated expressions of QCD spectral functions of XYZ-like states at leading order (LO) of perturbative QCD and up to dimensions $d\leq (6 - 8)$ non-perturbative condensates. 
The results are summarized in Tables \ref{tab:d-chl} to  \ref{tab:4q-su3}, from which we conclude, within the errors, that the observed XZ states are good candidates for being $1^{++}$ and $0^{++}$ molecules or/and four-quark states, contrary to the observed Y states which are too light compared to the predicted $1^{-\pm}$ and $0^{-\pm}$ states. We find that the SU3 breakings are relatively small for the masses ($\leq 10$ (resp. $3$)  $\%$) for the charm (resp. bottom) channels while they are large ($\leq 20\, \%$) for the couplings which decrease faster $(1/m_{b}^{3/2})$ than $1/m_{b}^{1/2}$ of HQET. QCD spectral sum rules (QSSR) approach cannot clearly separate (within the errors) molecules from four-quark states having the same quantum numbers.  Results for the $\bar BK (\bar DK)$-like molecules and $(\overline{Qq})(us)$-like four-quark states from\,\cite{SNX1} are also reviewed which do not favour  the molecule or/and four-quark interpretation of the $X(5568)$. We suggest to scan the charm $(2327 \sim 2444)~\mbox{MeV}$ and bottom $(5173 \sim 5226)~\mbox{MeV}$ regions for detecting the (unmixed)$(\overline{cu})ds$ and $(\overline{bu})ds$ via eventually their radiative or $\pi+$hadrons decays and reconsider more carefully the properties of the eventual $D^{*}_{s0}(2317)$ candidate. We expect that future experimental data and lattice results will check our predictions.
\end{abstract} 
%%%%%%%%%%%%%%%%%%%%%%%%%%%%%%%%%%%%%%%%%%%
\scriptsize
\end{frontmatter}
% \linenumbers
%\vspace*{-1cm}
%%%%%%%%%%%%%%%%%%%%%%%%%%%%%%%%%%%%%%%%%%%
\section{Introduction}
%%%%%%%%%%%%%%%%%%%%%%%%%%%%%%%%%%%%%%%%%%%
%\end{document}
%\vspace{-0.9cm}
%\nin
A large amount of exotic hadrons which differ from the standard $\bar{c}c$ charmonium and $\bar{b}b$ bottomium radial excitation states have been recently discovered in B-factories and have stimulated different theoretical interpretations. Most of them have been assigned as four-quark and/or molecule states. %\cite{MOLE1,MOLE2,MOLE3,MOLE4,MOLE5,MOLE5b,MOLE6,MOLE7,MOLE7b,MOLE8b,MOLE9,MOLE10,MOLE11,MOLE12}. 
In this paper, we use relativistic Laplace sum rules \cite{SVZa, BELL,BERT,NEUF,SNR}\footnote{For reviews, see \cite{SNB1, SNB2,SNB3,CK}} to improve some previous LO results for the masses and decay constants of the XYZ exotic heavy-light mesons obtained in the chiral limit \cite{X1A, X3B,NIELSEN}. In so doing, we include N2LO PT corrections to the heavy light correlators. We pursue our investigation by adding the SU3 NLO PT corrections. With these higher order (HO) PT contributions, we add the LO contribution of condensates up to dimension six. We do not include into the analysis contributions of condensates of higher dimension ($d\geq 8$) but only consider their effects as a source of systematic errors due to the truncation of the Operator Product Expansion (OPE). This work is a part of the original papers in \cite{SU3,SNCHI2,SNX2,SNX1,X3D}. 
%to the N2LO PT corrections in the chiral limit,
%%%%%%%%%%%%%%%%%%%%%%%%%%%%%%%%%%%%%%%%%%%
\section{Molecules and four-quark two point functions}
%%%%%%%%%%%%%%%%%%%%%%%%%%%%%%%%%%%%%%%%%%%
We shall work with the transverse part $\Pi^{(1)}_{mol}$ of the two-point spectral functions\,\footnote{Hereafter, similar expressions will be obtained for the four-quark states by replacing the sub-index {\it mol} by {\it 4q}.}:
\bea
\hspace*{-0.35cm}\Pi^{\mu\nu}_{mol}(q)\hspace{-0.3cm}&\equiv&\hspace{-0.3cm} i\int d^4x ~e^{iq.x}\lag 0
|T[{\cal O}_{mol}^{\mu}(x){\cal O}_{mol}^{\nu\dagger}(0)]
|0\rag\nnb\\
&=&\hspace*{-0.35cm} -\Pi^{(1)}_{mol}(q^2) \left(g^{\mu\nu}-\frac{q^\mu q^\nu}{q^2} \right)+\Pi^{(0)}_{mol}(q^2)\frac{q^\mu
q^\nu}{ q^2},
\lb{eq:2pf}
\eea
for the spin $1$ states while for the spin $0$ ones, we shall use the two-point functions $\psi_{mol}(q^2)$ built directly from the (pseudo)scalar currents:
\beq 
\psi_{mol}(q^2)=i\int d^4x ~e^{iq.x}\lag 0
|T[{\cal O}_{mol}(x){\cal O}_{mol}(0)]
|0\rag,
\label{eq:s02pf}
\eeq
which is related to $\Pi^{(0)}$ appearing in Eq.\,\ref{eq:2pf} via Ward identities \cite{SNB1, SNB2,BECCHI}.
%%%%%%%%
Thanks to their analyticity properties $\Pi^{(1,0)}_{mol}$ and $\psi_{mol}$ obey the dispersion relation:
\begin{align}
\Pi^{(1,0)}_{mol}(q^2) ;\psi_{mol}(q^2) &= \frac{1}{\pi}\int_{4 M^2_Q}^{\infty} \hspace*{-0.25cm}dt ~\frac{{\rm Im}\Pi^{(1,0)}_{mol}(t);{\rm Im}\psi_{mol}(t)}{t-q^2-i \epsilon},
\end{align}
where Im$\Pi^{(1,0)}_{mol}(t)$ and Im$\psi_{mol}(t)$ are the spectral functions.
%%%%%%%%%%
\subsection*{$\bullet$ Interpolating currents}
%%%%%%%%%%
\nin
The interpolating currents ${\cal O}_{mol}$ and ${\cal O}_{4q}$ respectively for the $\bar{D}D(\bar{B}B$)-like molecules and four-quark states  are given in Tables \ref{tab:curmol} and \ref{tab:cur4q}. The ones describing $\bar BK (\bar DK)$ (resp. $(\overline{Qq})(us)$)-like molecules (resp. four-quark) states are presented in Table \ref{tab:xcur}.
%%%%%%%%%%%%%%%%%%%%%%%%%%%%%%%%%%%%%%%%%%%
 {\scriptsize
\begin{table}[hbt]
%\begin{center}
%\begin{table*}[hbt]
\setlength{\tabcolsep}{0.7pc}
%\newlength{\digitwidth} \settowidth{\digitwidth}{\rm 0}
%\catcode`?=\active \def?{\kern\digitwidth}
 \caption{\footnotesize    
Interpolating currents ${\cal O}_{mol}$ with a definite C-parity describing the molecule-like states. $M\equiv D$(resp. $B$) and $Q\equiv c$(resp. $b$) for the $\bar{D}D$  (resp. $\bar{B}B$)-like molecules. $q\equiv d$(resp. $s$) for the chiral limit (resp. SU3 breaking).}
 {\scriptsize
\begin{tabular}{lll}
\hline
\hline
\\
States&$J^{PC}$&Molecule currents$\equiv {\cal O} _{mol}(x)$ \\
\\
\hline
%\\
{\bf Scalar}&$0^{++}$&\\
$\bar{M} M\, ,\, \bar{M}_sM_s$&&$(\bar{q}\gamma_5 Q)(\bar{Q}\gamma_5 q)$\\
$\bar{M}^* M^*\, ,\, \bar{M}^*_sM^*_s$&&$(\bar{q}\gamma_{\mu} Q)(\bar{Q}\gamma^{\mu} q)$\\
$\bar{M}^*_{0} M^*_{0}\, ,\, \bar{M}^*_{s0}M^*_{s0}$&&$(\bar{q}Q)(\bar{Q}q)$\\
$\bar{M}_{1} M_{1}\, ,\, \bar{M}_{s1}M_{s1}$&&$(\bar{q}\gamma_{\mu} \gamma_5 Q)(\bar{Q}\gamma^{\mu} \gamma_5 q)$\\
%\\
{\bf Axial-vector}&$1^{++}$&\\
$\bar{M}^* M\, ,\, \bar{M}^*_sM_s$&&$\frac{i}{\sqrt{2}}\left[(\bar{Q}\gamma_{\mu}q)(\bar{q}\gamma_5 Q)-(\bar{q}\gamma_{\mu} Q)(\bar{Q}\gamma_5 q)\right]$\\
$\bar{M}^*_{0} M_{1}\, ,\, \bar{M}^*_{s0}M_{s1}$&&$\frac{i}{\sqrt{2}}\left[(\bar{q}Q)(\bar{Q}\gamma_{\mu}\gamma_5 q)+(\bar{Q}q)(\bar{q}\gamma_{\mu}\gamma_5 Q)\right]$\\
%\\
\bf Pseudoscalar&$0^{- \pm}$& \\
$M^*_{0}M$,$M^*_{s0}M_s$&&$ \frac{1}{ \sqrt{2}}  \bigg[
		\big(  \bar{q}Q \big) \big( \bar{Q} \gamma_5 q \big) 
		\pm \:\big(  \bar{Q}q \big) \big( \bar{q} \gamma_5 Q \big)  \bigg]$\\
$M^*M_{1}$,$M^*_{s}M_{s1}$&&$ \frac{1}{ \sqrt{2}}  \bigg[
		\big(  \bar{Q}\gamma_{\mu}q \big) \big( \bar{q} \gamma^{\mu} \gamma_5 Q \big) 
		\mp \:\big(  \bar{Q}\gamma_{\mu} \gamma_5 q \big) \big( \bar{q} \gamma^{\mu} Q \big)  \bigg]$  \\
%\\
\bf Vector&$1^{- \pm}$&\\
$M^*_{0}M^*$,$M^*_{s0}M^*_s$&&$ \frac{1}{ \sqrt{2}}  \bigg[
		\big(  \bar{q}Q \big) \big( \bar{Q} \gamma_{\mu} q \big) 
		\mp \:\big(  \bar{Q}q \big) \big( \bar{q} \gamma_{\mu} Q \big)  \bigg] $  \\
$MM_{1}$,$M_{s}M_{s1}$&&$\frac{1}{ \sqrt{2}}  \bigg[
		\big(  \bar{Q} \gamma_{\mu} \gamma_5 q \big) \big( \bar{q}  \gamma_5 Q \big) 
		\pm \:\big(  \bar{q}\gamma_{\mu} \gamma_5 Q \big) \big( \bar{Q} \gamma_5 q \big)  \bigg] $   \\
%\\
\hline\hline
\end{tabular}
}
%\end{tabular*}
\label{tab:curmol}
%\end{center}
\end{table}
}
%%%%%%%%%%%%%%%%%%%%%%%%%%%%%%%%%%%%%%%%%%%
%%%%%%%%%%%%%%%%%%%%%%%%%%%%%%%%%%%%%%%%%%%
 {\scriptsize
\begin{table}[hbt]
\begin{center}
%\begin{table*}[hbt]
\setlength{\tabcolsep}{.2pc}
%\newlength{\digitwidth} \settowidth{\digitwidth}{\rm 0}
%\catcode`?=\active \def?{\kern\digitwidth}
 \caption{\footnotesize    
Interpolating currents describing the four-quark states. $Q\equiv c$ (resp. $b$) and $q\equiv d$(resp. $s$). $k$ is an arbitrary current mixing where the optimal value is found to be $k=0$ from \cite{X3B}}
 {\scriptsize
\begin{tabular}{lll}
\hline
\hline
\\
States&$J^{P}$&Four-quark currents$\equiv {\cal O} _{4q}(x)$ \\
\\
\hline
%\\
{\bf Scalar}&$0^{+}$&$\epsilon_{abc}\epsilon_{dec}\left[ (q^{T}_{a}C\gamma_5 Q_b)(\bar{q}_d \gamma_5 C \bar{Q}^{T}_{e})+k(q^{T}_{a} C Q_b)(\bar{q}_d  C \bar{Q}^{T}_{e})\right]$\\
%\\
{\bf Axial-vector}&$1^{+}$&$\epsilon_{abc}\epsilon_{dec}\left[ (q^{T}_{a}C\gamma_5 Q_b)(\bar{q}_d \gamma_{\mu} C \bar{Q}^{T}_{e})+k(q^{T}_{a} C Q_b)(\bar{q}_d  \gamma_{\mu} \gamma_5 C \bar{Q}^{T}_{e})\right]$\\
%\\
{\bf Pseudoscalar}&$0^{-}$&$\epsilon_{abc}\epsilon_{dec}\left[ (q^{T}_{a}C\gamma_5 Q_b)(\bar{q}_d C \bar{Q}^{T}_{e})+k(q^{T}_{a} C Q_b)(\bar{q}_d  \gamma_5 C \bar{Q}^{T}_{e})\right]$\\
%\\
{\bf Vector}&$1^{-}$&$\epsilon_{abc}\epsilon_{dec}\left[ (q^{T}_{a}C\gamma_5 Q_b)(\bar{q}_d \gamma_{\mu} \gamma_5 C \bar{Q}^{T}_{e})+k(q^{T}_{a} C Q_b)(\bar{q}_d  \gamma_{\mu} C \bar{Q}^{T}_{e})\right]$\\
%\\
\hline\hline
\end{tabular}
}
%\end{tabular*}
\label{tab:cur4q}
\end{center}
\vspace*{-0.5cm}
\end{table}
}
%%%%%%%%%%%%%%%%%%%%%%%%%%%%%%%%%%%%%%%%%%%
%%%%%%%%%%%%%%%%%%%%%%%%%%%%%%%%%%%%%%%%%%%
{\scriptsize
\begin{table}[H]
 \caption{\footnotesize Interpolating currents describing the $X(5568)$-like states. $M\equiv D$(resp. $B$) and $Q\equiv c$(resp. $b$).}  
%\tbl{
%}
\setlength{\tabcolsep}{1.1pc}
    {\scriptsize
 {\begin{tabular}{@{}lll@{}}% \toprule
&\\
\hline
\hline
\\
 Nature&$J^{P}$& Current   \\
\\
\hline
%\\
{\bf Molecule} &&\\
$\bar MK$&$0^{+}$&$(\bar Q\, i\gamma_5\ u)(\bar d \, i\gamma_5 \, s)$\\
$\bar M_s\pi$&$0^{+}$&$(\bar Q\, i\gamma_5\, s)(\bar d\, i\gamma_5\, u)$\\
$\bar M^*K$&$1^{+}$&$(\bar Q \,i\gamma_\mu\, u)(\bar d \,i\gamma_5\, s)$\\
$\bar M^*_s\pi$&$1^{+}$&$(\bar Q\, i\gamma_\mu s)(\bar d i\gamma_5\, u)$\\
%\\
{\bf Four-quark} &&\\
&$1^{-}$&$(s^T
C\gamma_5\,u)(\bar Q\,\gamma_\mu\gamma_5C\,\bar d^T)+k(s^T
C\,u)(\bar Q\,\gamma_\mu\, C\,\bar d^T)$\\
&$1^{+}$&$(s^T
C\gamma_5\,u)(\bar Q\,\gamma_\mu C\,\bar d^T)+k(s^T
C\,u)(\bar Q\,\gamma_\mu\gamma_5 C\,\bar d^T)$\\
%\\
\hline\hline
\end{tabular}}
\label{tab:xcur}
}
\end{table}
} 
%%%%%%%%%%
\subsection*{$\bullet$ Spectral function within MDA}
%%%%%%%%%%
\nin
We shall use the Minimal Duality Ansatz (MDA) for parametrizing the spectral function:
\beq
\frac{1}{\pi}\hspace{-0.1cm}\mbox{ Im}\Pi_{mol}(\hspace{-0.04cm} t \hspace{-0.03cm}) \hspace{-0.1cm}\simeq \hspace{-0.15cm}f^2_{H}M^8_{H}\delta(\hspace{-0.05cm}t-M_{H}^2)+\mbox{``QCD continuum"}\theta (\hspace{-0.04cm}t-t_c\hspace{-0.05cm}),
\label{eq:mda}
\eeq
where $f_H$ is the decay constant defined as:
\beq
\la 0| {\cal O}_{mol}|H\ra\hspace{-0.08cm}=\hspace{-0.08cm}f_{H}M^4_{H}~,~\la 0| {\cal O}_{mol}^\mu|H\ra\hspace{-0.08cm}=\hspace{-0.08cm}f_{H}M^5_{H}\epsilon_\mu,
\label{eq:coupling}
\eeq
respectively for spin 0 and 1 states with  $\epsilon_\mu$ the (axial-)\newline vector polarization. Noting that in the previous definition in Tables \ref{tab:curmol}--\ref{tab:xcur},  the bilinear (pseudo)scalar current acquires an anomalous dimension due to its normalization, thus the decay constants run  to order $\alpha_s^2$ as\,\footnote{The coupling of the (pseudo)scalar molecule built from two (axial)-vector currents has no anomalous dimension and does not run.}:
\bea
f^{(s,p)}_{mol}(\mu)&=&\hat f^{(s,p)}_{mol} \ga -\beta_1a_s\dr^{4/\beta_1}/r_m^2~,\nnb \\
f^{(1)}_{mol}(\mu)&=&\hat f^{(1)}_{mol} \ga  -\beta_1a_s\dr^{2/\beta_1}/r_m~,
\label{eq:fhat}
\eea
where we have introduced the renormalization group invariant coupling $\hat f_{mol}$; 
%$\mu$ is an arbitrary  subtraction constant; 
$-\beta_1=(1/2)(11-2n_f/3)$ is the first coefficient of the QCD $\beta$-function for $n_f$ flavours and  $a_s\equiv (\alpha_s/\pi)$.  The QCD corrections numerically read;
\bea
r_m(n_f=4)=1+1.014 a_s +1.389a_s^2~,\nnb \\
r_m(n_f=5)=1+1.176a_s +1.501a_s^2.  
\eea
The higher order states contributions are smeared by the ``QCD continuum" coming from the discontinuity of the QCD diagrams and starting from a constant threshold $\sqrt{t_c}$.
%%%%%%%%%%
\subsection*{$\bullet$ NLO and N2LO PT corrections using factorization}
%%%%%%%%%%
Assuming a factorization of the four-quark interpolating current as a natural consequence of the molecule definition of the state, we can write the corresponding spectral function as a convolution of the ones associated to quark bilinear current. In this way, we obtain \cite{PICH} for the $\bar{D}D^*$ and $\bar{D}^*_0 D^*$-like spin 1 states:
\bea
\frac{1}{ \pi}{\rm Im} \Pi^{(1)}_{mol}(t)\hspace{-0.3cm}&=&\hspace{-0.3cm}\theta (t-4M_Q^2) \hspace{-0.05cm} \ga \hspace{-0.05cm}\frac{1}{ 4\pi}\hspace{-0.05cm}\dr^2 \hspace{-0.12cm} t^2\hspace{-0.2cm} \int_{M_Q^2}^{(\sqrt{t}-M_Q)^2} \hspace*{-1cm}dt_1 \hspace{0.25cm}\int_{M_Q^2}^{(\sqrt{t}-\sqrt{t_1})^2} \hspace*{-1cm}dt_2\nnb\\
&&\hspace{-0.3cm}\times\lambda^{3/2}\frac{1}{ \pi}{\rm Im} \Pi^{(1)}(t_1) \frac{1}{ \pi}{\rm Im} \psi^{(s,p)}(t_2)~.
\label{eq:convolution}
\eea
For the $\bar{D}D$ spin $0$ state, one has:
\bea
\frac{1}{ \pi}{\rm Im} \psi^{(s)}_{mol}(t)\hspace{-0.08cm}=\hspace{-0.08cm}\theta (t-4M_Q^2) \hspace{-0.05cm} \ga \hspace{-0.05cm}\frac{1}{ 4\pi}\hspace{-0.05cm}\dr^2 \hspace{-0.12cm} t^2\hspace{-0.2cm} \int_{M_Q^2}^{(\sqrt{t}-M_Q)^2} \hspace*{-1cm}dt_1 \hspace{0.25cm}\int_{M_Q^2}^{(\sqrt{t}-\sqrt{t_1})^2} \hspace*{-1.1cm}dt_2 \nnb\\
\times\lambda^{1/2} \ga  \frac{t_1}{ t}\hspace{-0.1cm}+\frac{t_2}{ t}-1  \dr^2\frac{1}{ \pi}{\rm Im}\psi^{(p)}(t_1) \frac{1}{ \pi} {\rm Im} \psi^{(p)}(t_2),
\eea

and for the $\bar{D}^*D^*$ spin $0$ state:
\bea
\frac{1}{ \pi}{\rm Im} \psi^{(s)}_{mol}(t)\hspace{-0.08cm}=\hspace{-0.08cm}\theta (t-4M_Q^2) \hspace{-0.05cm} \ga \hspace{-0.05cm}\frac{1}{ 4\pi}\hspace{-0.05cm}\dr^2 \hspace{-0.12cm} t^2\hspace{-0.2cm} \int_{M_Q^2}^{(\sqrt{t}-M_Q)^2} \hspace*{-1cm}dt_1 \hspace{0.25cm}\int_{M_Q^2}^{(\sqrt{t}-\sqrt{t_1})^2} \hspace*{-1cm}dt_2\nnb\\\times\lambda^{1/2}\times\left[\ga\frac{t_1}{ t}+\frac{t_2}{ t}-1\dr^2+\frac{8t_1 t_2}{t^2}\right]\nnb\\
\times\frac{1}{ \pi}{\rm Im}\Pi^{(1)}(t_1) \frac{1}{ \pi} {\rm Im} \Pi^{(1)}(t_2),
\eea

where:
\beq
\lambda=\ga 1-\frac{\ga \sqrt{t_1}- \sqrt{t_2}\dr^2}{ t}\dr \ga 1-\frac{\ga \sqrt{t_1}+ \sqrt{t_2}\dr^2}{ t}\dr~,
\eeq
is the phase space factor and $M_Q$ is the on-shell heavy quark mass. ${\rm Im}\Pi^{(1)}(t) $ is the spectral function associated to the bilinear $\bar{c}\gamma_{\mu}(\gamma_5)q$ (axial-)vector current, while ${\rm Im}\,\psi^{(s,p)}(t) $ is associated to the $\bar{c}i(\gamma_5)q$ (pseudo)scalar current\footnote{In the chiral limit $m_q=0$, the PT expressions of the vector (resp. scalar) and axial-vector (resp. pseudoscalar) spectral functions are the same.}. We shall assume that a such factorization also holds for four-quark states. 
%%%%%%%%%%
\subsection*{$\bullet$ The inverse Laplace transform sum rule (LSR)}
%%%%%%%%%%
The LSR and its ratio read:
\beq
{\cal L}_{mol}(\tau,t_c,\mu)=\frac{1}{\pi}\!\int_{4M_Q^2}^{t_c}\!dt\,e^{-t\tau}\mbox{Im}\{\Pi_{mol} ; \psi_{mol}\}(t,\mu),
\label{eq:LSR}
\eeq
\beq
{\cal R}_{mol}(\tau,t_c,\mu)\! =\! \frac{\int_{4M_Q^2}^{t_c}\! dt\,t\,e^{-t\tau}\mbox{Im}\{\Pi_{mol} ; \psi_{mol}\}(t,\mu)}
{\int_{4M_Q^2}^{t_c}\! dt\, e^{-t\tau} \mbox{Im}\{\Pi_{mol} ; \psi_{mol}\}(t,\mu)}\!\simeq \! M_R^2,
\eeq
where $\mu$ is the subtraction point which appears in the approximate QCD series when radiative corrections are included and $\tau$ is the sum rule variable replacing $q^2$.
%%%%%%%%%%
\subsection*{$\bullet$ Double ratios of inverse Laplace transform sum rule}
%%%%%%%%%%
Double Ratios of Sum Rules (DRSR)\,\cite{DRSR,SNmassa,SNmassb,SNFORM1,SNGh3,SNGH,SNhl,HBARYON1,HBARYON2,NAVARRA,SNB1,SNB2} 
are also useful for extracting the SU3 breaking effects on couplings and mass ratios. They read:
\beq
f^{sd}_{mol}\equiv \frac{{\cal L}_{mol}^{s}(\tau,t_c,\mu)}{{\cal L}_{mol}^{d}(\tau,t_c,\mu)}~, ~~~r_{mol}^{sd}\equiv \frac{{\cal R}_{mol}^{s}(\tau,t_c,\mu)}{{\cal R}_{mol}^{d}(\tau,t_c,\mu)},
\label{eq:DRSR}
\eeq
where the upper indices $s,d$ indicates the $s$ and $d$ quark channels. These DRSR can be used when each sum rule optimizes at the same values of the parameters $(\tau,t_c,\mu)$.

%%%%%%%%%%
\subsection*{$\bullet$ Stability criteria and some phenomenological tests}
%%%%%%%%%%
The variables $\tau,\mu$ and $t_c$ are, in principle, free external parameters. We shall use stability criteria (if any) with respect to these free 3 parameters, for extracting the optimal results. In the standard MDA given in Eq.\,\ref{eq:mda} for parametrizing the spectral function, the ``QCD continuum" threshold $\sqrt{t_c}$ is constant and is independent on the subtraction point $\mu$. One should notice that this standard MDA with constant $\sqrt{t_c}$ describes quite well the properties of the lowest ground state as explicitly demonstrate in \cite{SNFB12a} and in various examples \cite{SNB1, SNB2} after confronting the integrated spectral function within this simple parametrization with the full data measurements. It has been also successfully tested in the large $N_c$ limit of QCD in \cite{PERISb}. Though it is difficult to estimate with a good precision the systematic error related to this simple model for reproducing accurately the data, we expect that the same feature is reproduced for the case of the XYZ discussed here where complete data are still lacking.
%%%%%%%%%%%%%%%%%%%%%%%%%%%%%%%%%%%%%%%%%%%
\section{QCD input parameters}
%%%%%%%%%%%%%%%%%%%%%%%%%%%%%%%%%%%%%%%%%%%
The QCD parameters which shall appear in the following analysis will be the charm and bottom quark masses $m_{c,b}$, the strange quark mass $m_s$ (we shall neglect  the light quark masses $m_{u,d}$), the light quark condensate $\qq$ ($q\equiv u,d$),  the gluon condensates $ \lag\alpha_sG^2\rag \equiv \la \alpha_s G^a_{\mu\nu}G_a^{\mu\nu}\ra$ 
and $ \la g^3G^3\ra \equiv \la g^3f_{abc}G^a_{\mu\nu}G^b_{\nu\rho}G^c_{\rho\mu}\ra$, 
the mixed condensate $\la\bar qGq\ra \equiv {\la\bar qg\sigma^{\mu\nu} (\lambda_a/2) G^a_{\mu\nu}q\ra}=M_0^2\la \bar qq\ra$ 
and the four-quark  condensate $\rho\alpha_s\la\bar qq\ra^2$, where  $\rho\simeq 3-4$ indicates the deviation from the four-quark vacuum saturation. Their values are given in Table \ref{tab:param} and more recently in\,\cite{SN18}. The original errors on 
$\kappa\equiv \la \bar ss\ra/\la\bar dd\ra$ have been enlarged to take into account the lattice result\,\cite{MCNEILE} which needs to be checked by some other groups. 
We shall work with the running light quark condensates, which read to leading order in $\alpha_s$: 
 \bea
{\la\bar qq\ra}(\tau)&=&-{\hat \mu_q^3  \ga-\beta_1a_s\dr^{2/{\beta_1}}},\nnb\\
{\la\bar q Gq\ra}(\tau)&=&-{M_0^2{\hat \mu_q^3} \ga-\beta_1a_s\dr^{1/{3\beta_1}}},
\label{d4g}
\eea
and the running quark mass to NLO (for the number of flavours $n_f=3$)
\beq
\overline{m}_s(\tau)=\hat m_s(-\beta_1 a_s)^{-2/\beta_1}(1+0.8951a_s),
\eeq
where %$\beta_1=-(1/2)(11-2n_f/3)$ is the first coefficient of the $\beta$ function for $n_f$ flavours; $a_s\equiv \alpha_s(\tau)/\pi$; 
$\hat\mu_q$ and $\hat m_s$ is the spontaneous RGI light quark condensate \cite{FNR} and strange quark mass.
%%%%%%%%%%%%%%%%%%%%%%%%%%%%%%%%%%%%%%%%%%%
%%%%%%%%%%%%%%%%%%%%%%%%%%%%%%%%%%%%%%%%%%%
{\scriptsize
\begin{table}[hbt]
\setlength{\tabcolsep}{.2pc}
 \caption{QCD input parameters:the original errors for $\la\alpha_s G^2\ra$, $\la g^3  G^3\ra$ and $\rho \la \bar qq\ra^2$ have been multiplied by about a factor 3 for a conservative estimate of the errors (see also the text). }  
%\tbl{
%}
    {\footnotesize
 % \begin{tabular}{lll}
 {\begin{tabular}{@{}lll@{}}
&\\
\hline
\hline
\\
Parameters&Values& Ref.    \\
\\
\hline
%\\
$\alpha_s(M_\tau)$& $0.325(8)$&\cite{SNTAU,BNPa,BNPb,BETHKE}\\
%,PICHTAU,BETHKE,PDG}\\
$\hat m_s$&$(0.114\pm0.006)$ GeV &\cite{SNB1,SNTAU,SNmassa,SNmassb,SNmass98a,SNmass98b,SNLIGHT}\\
$\overline{m}_c(m_c)$&$1261(12)$ MeV &average \cite{SNmass02,SNH10a,SNH10b,SNH10c,PDG,IOFFEa,IOFFEb}\\
$\overline{m}_b(m_b)$&$4177(11)$ MeV&average \cite{SNmass02,SNH10a,SNH10b,SNH10c,PDG}\\
$\hat \mu_q$&$(253\pm 6)$ MeV&\cite{SNB1,SNmassa,SNmassb,SNmass98a,SNmass98b,SNLIGHT}\\
$\kappa\equiv \la \bar ss\ra/\la\bar dd\ra$& $(0.74^{+0.34}_{- 0.12})$&\cite{HBARYON1,HBARYON2,SNB1}\\
%$\la \bar dd\ra(2) $&$-(275.7\pm 6.6)^3$ MeV$^3$&\cite{SNB1,SNmass}\\
$M_0^2$&$(0.8 \pm 0.2)$ GeV$^2$&\cite{JAMI2a,JAMI2b,JAMI2c,HEIDb,HEIDc,SNhl}\\
$\la\alpha_s G^2\ra$& $(7\pm 3)\times 10^{-2}$ GeV$^4$&
\cite{SNTAU,LNT,SNIa,SNIb,YNDU,SNH10a,SNH10b,SNH10c,SNG2,SNGH}\\
$\la g^3  G^3\ra$& $(8.2\pm 2.0)$ GeV$^2\times\la\alpha_s G^2\ra$&
\cite{SNH10a,SNH10b,SNH10c}\\
$\rho \alpha_s\la \bar qq\ra^2$&$(5.8\pm 1.8)\times 10^{-4}$ GeV$^6$&\cite{SNTAU,LNT,JAMI2a,JAMI2b,JAMI2c}\\
%$\hat m_s$&$(0.114\pm0.006)$ GeV &\cite{SNB1,SNTAU9,SNmassa,SNmassb,SNmass98a,SNmass98b,SNLIGHT}\\
%$\kappa\equiv \la \bar ss\ra/\la\bar dd\ra$& $(0.74^{+0.34}_{- 0.12})$&\cite{HBARYONa,HBARYONb,SNB1}\\
%\\
\hline\hline
\end{tabular}}
}
\label{tab:param}
\vspace*{-0.5cm}
%\caption{%\scriptsize   
\end{table}
} 
%%%%%%%%%%%%%%%%%%%%%%%%%%%%%%%%%%%%%%%%%%%
\section{QCD expressions of the spectral functions}
%%%%%%%%%%%%%%%%%%%%%%%%%%%%%%%%%%%%%%%%%%%
%\vspace*{-1cm}
In our works \cite{SNX1,SNX2,SU3}, we provide new compact integrated expressions of QCD spectral functions at LO of PT QCD and including non-perturbative (NP) condensates having dimensions $d\leq 6-8$. NLO and N2LO corrections are introduced using the convolution integrals in Eq.\,\ref{eq:convolution}. The expressions of the QCD spectral functions of heavy-light bilinear currents are known to order $\alpha_s$ (NLO) from\, \cite{BROAD} and to order $\alpha^2_s$ (N2LO) in the chiral limit $m_q=0$ from\,\cite{CHETa,CHETb} which are available as a Mathematica program named Rvs. We shall use the SU3 breaking PT corrections at NLO\, \cite {GELH} from the two-point function formed by bilinear currents. N3LO corrections are estimated from a geometric growth of the QCD PT series\,\cite{SNZ} as a source of PT errors, which we expect to give a good approximation of the uncalculated higher order terms dual to the $1/q^2$ contribution of a tachyonic gluon mass\,\cite{CNZ1, CNZ2}.

In our analysis, we replace the on-shell (pole) mass appearing in the spectral functions with the running mass using the relation, to order $\alpha^2_s$\,\cite{SNB1,SNB2,SNB3}:
\bea
M_Q \hspace{-0.3cm}&=&\hspace{-0.3cm} \overline{m}_Q(\mu)\Big{[}1+\frac{4}{3} a_s+ (16.2163 -1.0414 n_l)a_s^2\nnb\\
&&\hspace{-0.3cm}+\ln{\ga\frac{\mu}{ M_Q}\dr^2} \ga a_s+(8.8472 -0.3611 n_l) a_s^2\dr\nnb\\
&&\hspace{-0.3cm}+\ln^2{\ga\frac{\mu}{ M_Q}\dr^2} \ga 1.7917 -0.0833 n_l\dr a_s^2...\Big{]},
\label{eq:pole}
\eea
for $n_l$ light flavours where $\mu$ is the arbitrary subtraction point.% and $a_s\equiv \alpha_s / \pi$.
%%%%%%%%%%%%%%%%%%%%%%%%%%%%%%%%%%%%%%%%%%%
\section{Tests of the Factorization Assumption}
%%%%%%%%%%%%%%%%%%%%%%%%%%%%%%%%%%%%%%%%%%%
%%%%%%%%%%
\subsection*{$\bullet$ Factorization test for PT$\oplus$NP contributions at LO}
%%%%%%%%%%
\begin{figure}[hbt] 
\begin{center}
{\includegraphics[width=5cm  ]{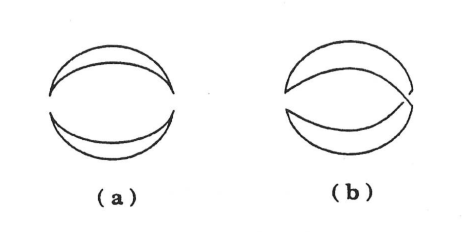}}
\caption{
\scriptsize 
{\bf (a)} Factorized contribution to the four-quark correlator at lowest order of PT; {\bf (b)} Non-factorized contribution at lowest order of PT (the figure comes from\, \cite{PICH}).
}
\label{fig:factor} 
\end{center}
\end{figure} 
From our previous work\,\cite{SNX2,SNCHI2}, we have noticed that assuming a factorization of the PT at LO and including NP contributions induces an effect about $2.2 \%$ for the decay constant and $0.5 \%$ for the mass, which is quite tiny. However, we shall work in the following with the full non-factorized PT$\oplus$NP of the LO expressions.
%%%%%%%%%%
\subsection*{$\bullet$ Test at NLO of PT from the $\bar{B}^0B^0$ four-quark correlator }
%%%%%%%%%%
\begin{figure}[hbt] 
\begin{center}
{\includegraphics[width=6cm  ]{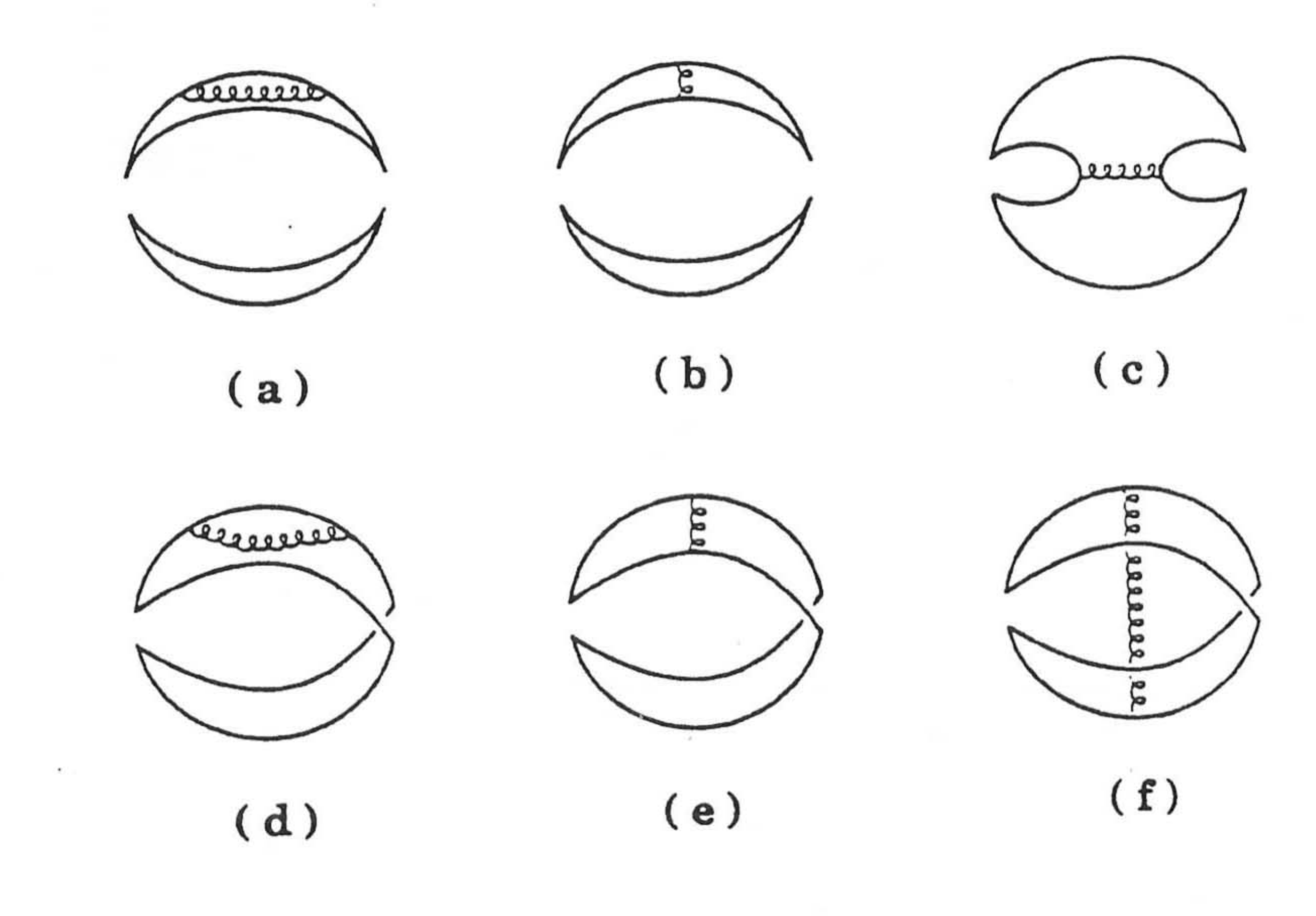}}
\caption{
\scriptsize 
{\bf (a,b)} Factorized contribution to the four-quark correlator at NLO of PT; {\bf (c to f)} Non-factorized contribution at NLO of PT (the figure comes from\, \cite{PICH}).
}
\label{fig:factoras} 
\end{center}
\end{figure} 
For extracting the PT $\alpha^n_s$ corrections to the correlator and due to the technical complexity of the calculations, we shall assume that these radiative corrections are dominated by the ones from factorized diagrams while we neglect the ones from non-factorized diagrams. This fact has been proven explicitly by\, \cite{BBAR2, BBAR3} in the case of $\bar{B}^0B^0$ systems (very similar correlator as the ones discussed in the following) where the non-factorized $\alpha_s$ corrections do not exceed $10 \%$ of the total $\alpha_s$ contributions
%%%%%%%%%%
\subsection*{$\bullet$ Conclusions of the factorization tests}
%%%%%%%%%%
We expect from the previous LO examples that the masses of the molecules are known with a good accuracy while, for the coupling, we shall have in mind the systematics induced by the radiative corrections estimated by keeping only the factorized diagrams where their contributions will be extracted from the convolution integrals given in Eq.\,5. Here, the suppression of the NLO corrections will be more pronounced for the extraction of the meson masses from the ratio of sum rules compared to the case of the $\bar{B}^0B^0$ systems.
%%%%%%%%%%%%%%%%%%%%%%%%%%%%%%%%%%%%%%%%%%%
\section{Molecules and four-quark states}
%%%%%%%%%%%%%%%%%%%%%%%%%%%%%%%%%%%%%%%%%%%
We shall study the charm channels and their beauty analogue in chiral limit (resp. SU3 breaking). As the analysis will be performed using the same techniques, we shall illustrate it in the case of $\bar{D}D$ (resp. $\bar{D}_s D_s$). The results are given in Tables \ref{tab:d-chl} to \ref{tab:4q-chl} (resp. Tables \ref{tab:d-su3} to \ref{tab:4q-su3}) for the chiral limit (resp. SU3 breaking) case.
%%%%%%%%%%
\subsection*{$\bullet$ $\bar{D}D$ state in chiral limit}
\begin{figure}[hbt] 
\begin{center}
{\includegraphics[width=3.8cm]{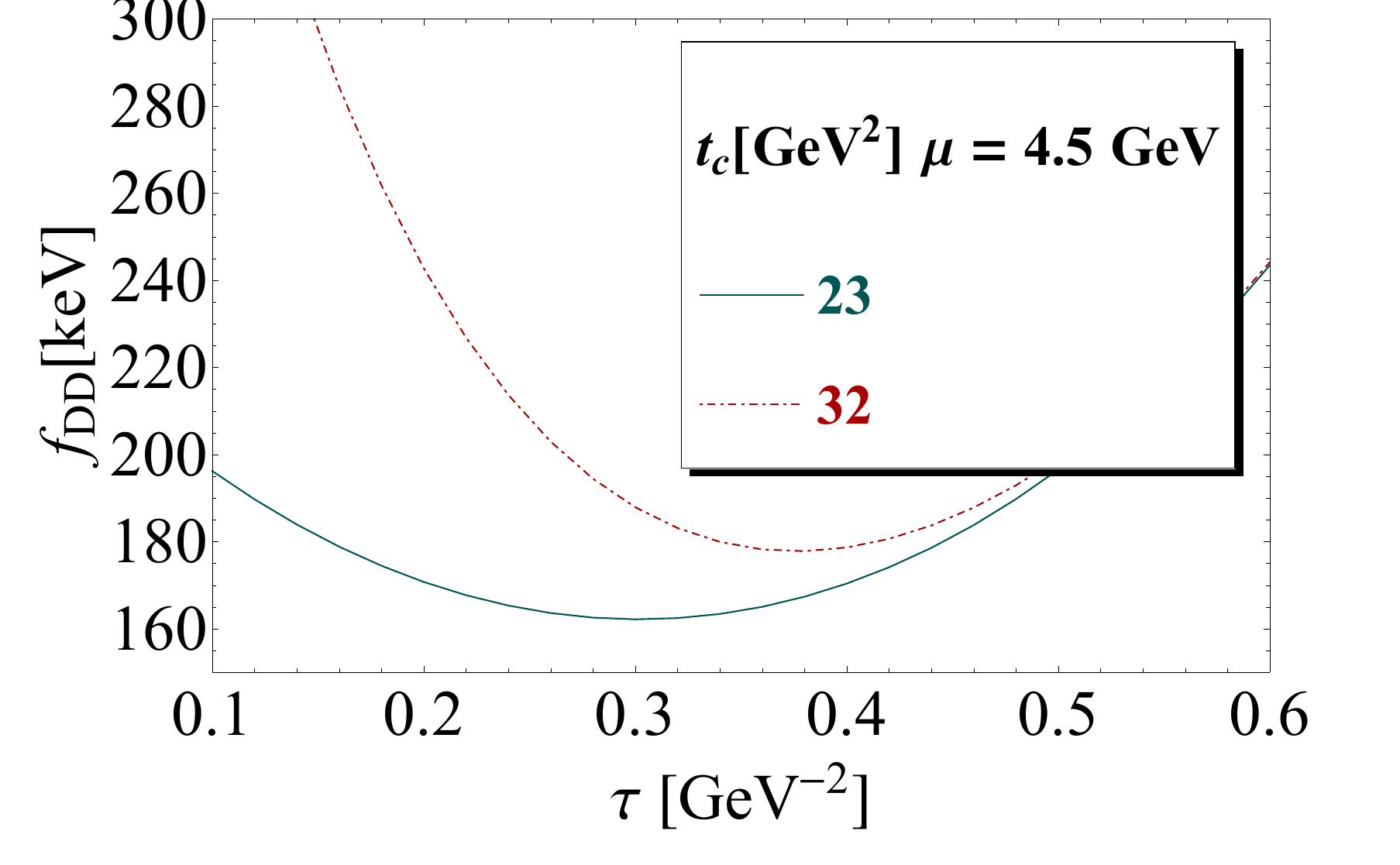}}
{\includegraphics[width=3.8cm]{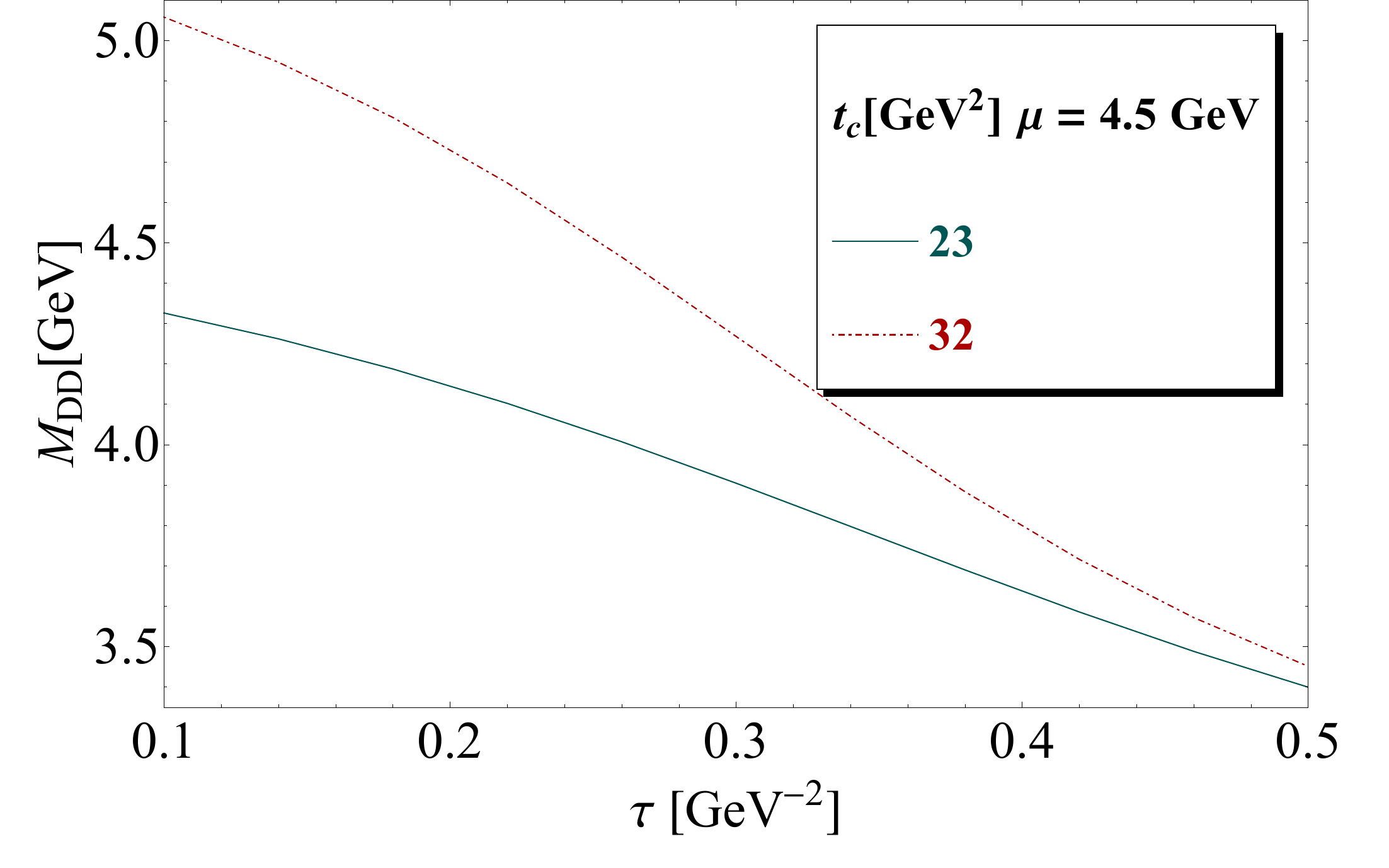}}
\centerline {\hspace*{-0.5cm} a)\hspace*{3cm} b) }
\caption{
\scriptsize 
{\bf a)} The coupling $f_{DD}$  at N2LO as function of $\tau$ for different values of $t_c$, for $\mu=4.5$ GeV  and for the QCD parameters in Table\,\ref{tab:param}; {\bf b)} The same as a) but for the mass $M_{DD}$.
}
\label{fig:dd-n2lo} 
\end{center}
\end{figure}

%%%%%%%%%%
We study the behavior of the coupling\footnote{Here and in the following: decay constant is the same as coupling} and mass in term of LSR variable $\tau$ for different values of $t_c$ at N2LO as shown in Fig.\,\ref{fig:dd-n2lo}.
%%%%%%%%%%%%%%%%%%%%%%%%%
\nin
%%%%%%%%%%%%%%%%%%%%%%%%%
 We consider as final and conservative results the one corresponding to the beginning of the $\tau$ stability for $t_c=23 ~\mbox{GeV}^2$ and $\tau\simeq 0.25 ~\mbox{GeV}^{-2}$ until the one where $t_c$ stability is reached for $t_c\simeq 32 ~\mbox{GeV}^2$ and $\tau\simeq 0.35 ~\mbox{GeV}^{-2}$. 
%%%%%%%%%%
\subsubsection*{-- Convergence of the PT series} 
\begin{figure}[hbt] 
\begin{center}
{\includegraphics[width=3.8cm]{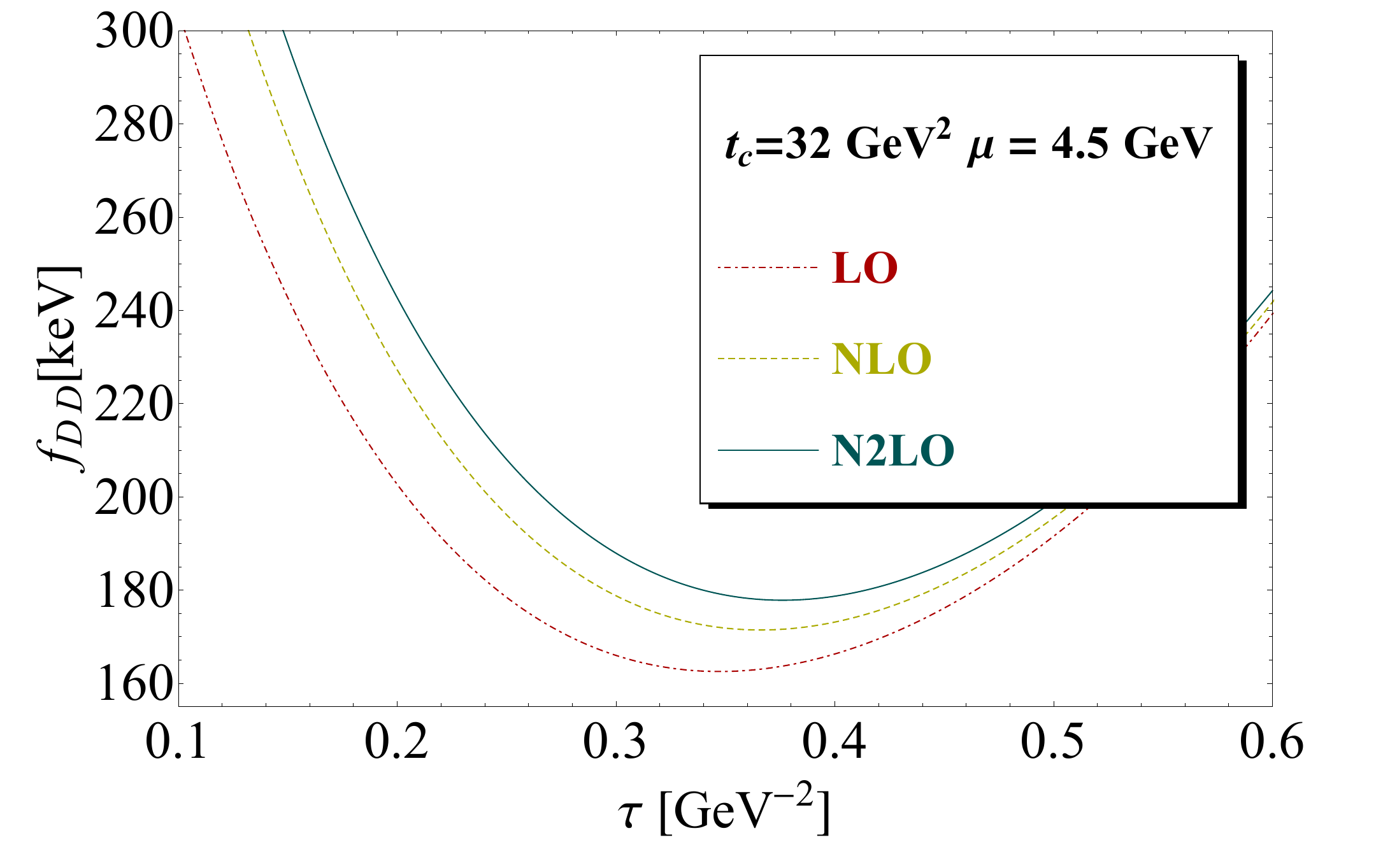}}
{\includegraphics[width=3.8cm]{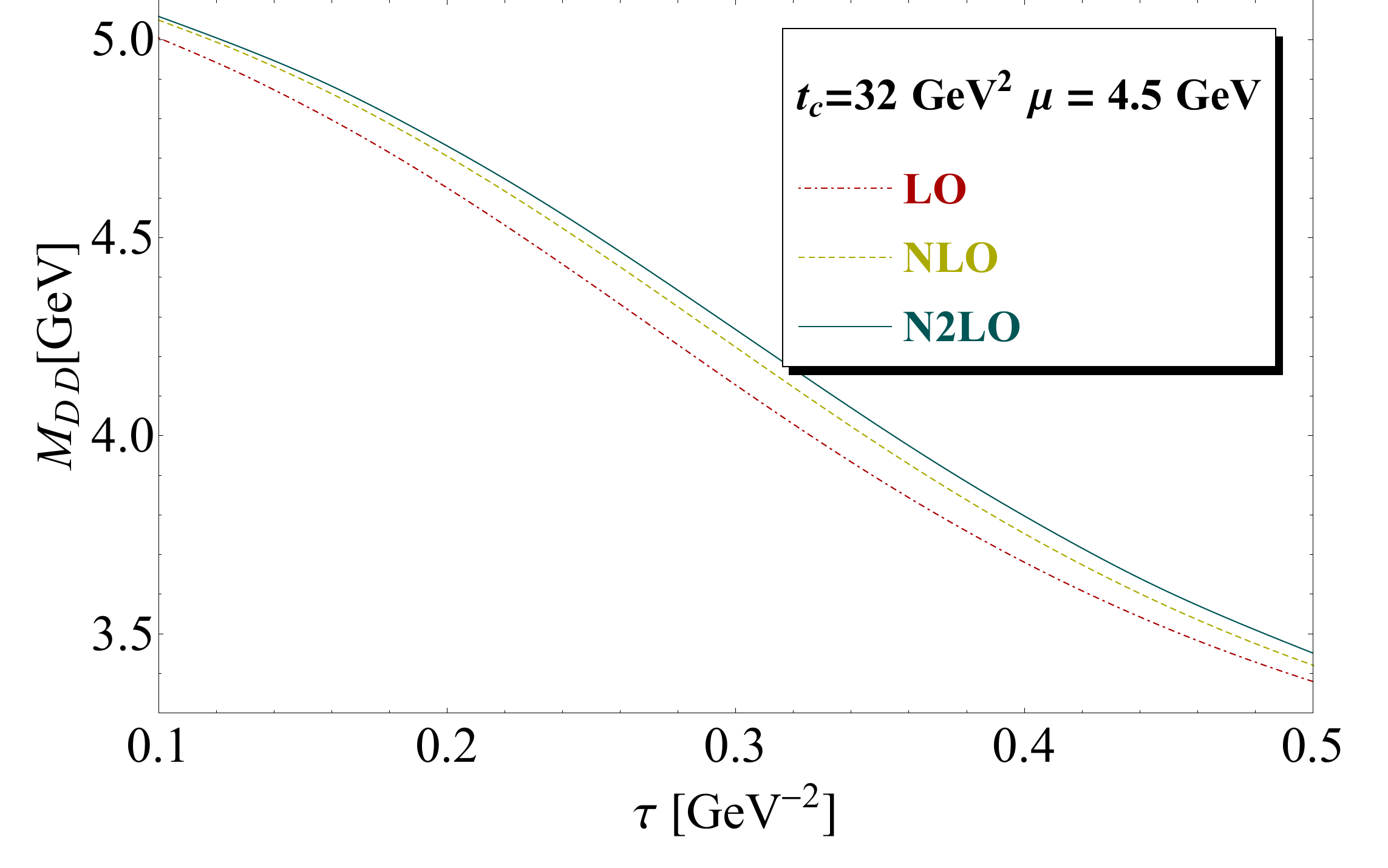}}
\centerline {\hspace*{-0.5cm} a)\hspace*{3cm} b) }
\caption{
\scriptsize 
{\bf a)} $\tau$-behavior of $f_{\bar DD}$ for $t_c=32 ~\mbox{GeV}^2$ and $\mu=4.5 ~\mbox{GeV}$ and for different truncation of the PT series; {\bf b)} the same as {\bf a)} but for $M_{\bar DD}$.
}
\label{fig:dd-pt} 
\end{center}
\end{figure}

%%%%%%%%%%
\nin
Using the previous value of $t_c\simeq 32 ~\mbox{GeV}^2$, we study in Fig.\,{\ref{fig:dd-pt}} the convergence of the PT series for a given value of $\mu=4.5 ~\mbox{GeV}$. We observe that from NLO to N2LO, the mass decreases by $\sim 0.1\%$. This result indicates a good convergence  of the PT series.
%\cite{RAPHAEL}.
%%%%%%%%%%%%%%%%%%%%%%%%%
\nin
%%%%%%%%%%
\subsubsection*{-- $\mu$-stability}
\begin{figure}[hbt] 
\begin{center}
{\includegraphics[width=3.8cm]{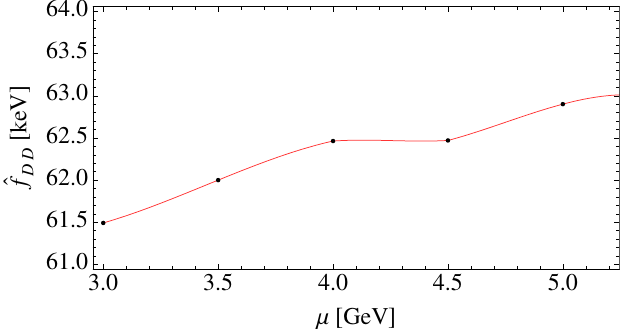}}
{\includegraphics[width=3.8cm]{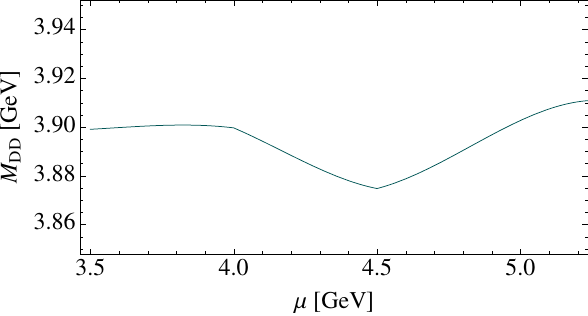}}
\centerline {\hspace*{-0.5cm} a)\hspace*{3cm} b) }
\caption{
\scriptsize 
{\bf a)} $\mu$-behavior of $\hat{f}_{\bar DD}$ at N2LO; {\bf b)} $\mu$-behavior of $M_{\bar DD}$ at N2LO.
}
\label{fig:dd-mu} 
\end{center}
\end{figure}

%%%%%%%%%%
\nin
We improve our previous results by using different values of $\mu$ (Fig.\,{\ref{fig:dd-mu}}). Using the fact that the final results must be independent of the arbitrary parameter $\mu$, we consider as optimal results the one at the inflexion point
for $\mu\simeq (4.0-4.5)$ GeV.
%%%%%%%%%%%%%%%%%%%%%%%%%
%\bea
%M_{\bar DD}&=&3898(36) \rm{MeV}~,\nnb \\
%\hat{f}_{\bar DD}&=&(62\pm 6)~{\rm keV}\nnb\\
%\lrar f_{\bar DD}(4.5)&=&(170\pm 15)~{\rm keV}.
%\label{res:dd}
%\eea
%%%%%%%%%%%%%%%%%%%%%%%%%
\nin
%%%%%%%%%%%%%%%%%%%%%%%%%%%%%%%%%%%%%%%
%%%%%%%%%%%%%%%%%%%%%%%%%%%%%%%%%%%%%%%
%%%   RESULTS CHIRAL LIMIT         %%%%
%%%%%%%%%%%%%%%%%%%%%%%%%%%%%%%%%%%%%%%
%%%%%%%%%%%%%%%%%%%%%%%%%%%%%%%%%%%%%%%
{\scriptsize
\begin{table*}[hbt]
\setlength{\tabcolsep}{0.8pc}
 \caption{$\bar DD$-like molecules masses, invariant and running couplings  from LSR within stability criteria at LO to N2LO of PT. The invariant coupling $\hat f_X$ is defined in Eq.\,\ref{eq:fhat}.}  
% space before first and after last column: 1.5pc
% space between columns: 3.0pc (twice the above)
% -----------------------------------------------------
% adapted from TeX book, p. 241
%\newlength{\digitwidth} \settowidth{\digitwidth}{\rm 0}
%\catcode`?=\active \def?{\kern\digitwidth}
% -----------------------------------------------------
{\scriptsize{
%\begin{tabular}{\textwidth}{@{}l@{\extracolsep{\fill}}  lll   lll  lll  c}
\begin{tabular*}{\textwidth}{@{}lll   lll  lll  l cc@{\extracolsep{\fill}}l}
%\begin{tabular}{lll   lll  lll  l c}
\\
\hline
\hline
%\\
                \bf Nature& \multicolumn{3}{c}{${\hat f_X}$  [keV]} 
                 & \multicolumn{3}{c }{${ f_X(4.5)}$  [keV]} 
                 &  \multicolumn{3}{c}{ Mass  [MeV]} 
                  & Threshold
                 &  Exp.
                 \\
\cline{2-4} \cline{5-7}\cline{8-10}
%\\
                 & \multicolumn{1}{l}{LO} 
                 & \multicolumn{1}{l}{NLO} 
                 & \multicolumn{1}{l }{ N2LO} 
                      & \multicolumn{1}{l}{LO} 
                 & \multicolumn{1}{l}{NLO} 
                 & \multicolumn{1}{l }{ N2LO} 
                   & \multicolumn{1}{l}{LO} 
                 & \multicolumn{1}{l}{NLO} 
                 & \multicolumn{1}{l}{ N2LO} 
                  \\
%\\
\hline
%\\
 $\bf {J^{PC}=0^{++}}$ &&&&&&&&&&&--\\
$\bar DD$&56&60& 62(6)&155&164& 170(15)&3901&3901& 3898(36)&3739\\
$\bar D^*D^*$&--&--&--&269&288&302(47)&3901&3903&3903(179)&4020&\\
$ D^*_0D^*_0$&--&--&--&--&97&114(18)&--&4003& 3954(223)&4636\\
$\bar D_{1}D_{1}$&--&--&--&--&236&274(37)&--&3838&3784(56)&-- \\
\\
$\bf {J^{PC}=1^{+\pm}} $ &&&&&&&&&&&$X_c,Z_c$\\
%(3872)$, \\
%\,\cite{BELLEX,BABARX1,BABARX2,CDF,D0,LHCbX
%&&&&&&&&&&&$Z_c(3900,4025,4200,4430)$\\
%\,\cite{BELLEZ1,BELLEZ2,BES}, 
%$Z_c(4430)$
%\,\cite{BELLEZ3,LHCbZ}\\
$\bar D^*D$&87&93&97(10)&146&154&161(17)&3901&3901&3903(62)&3880\\
$\bar D^*_0D_1$&--&--&--&--&96&112(17)&--&3849&3854(182)&4739\\
\\
%%%%%%%%%%%%%%%%%%%%%%%%%%%%%%%%%%%%%%%%%%%%%%%%%%%
$\bf  {J^{PC}=0^{-\pm}} $ &&&&&&&&&&&--\\
$\bar D^*_0D$&68&88&94(7)&190&240&257(19)&5956&5800&5690(140)&4188\\
$\bar D^*D_1$&--&--&--&382&490&564( 38)&6039&5898&5797(141)&4432&\\
\\
%%%%%%%%%%%%%%%%%%%%%%%%%%%%%%%%%%%%%%%%%%%%%%%%%%
$\bf { J^{PC}=1^{--}} $ &&&&&&&&&&&$Y_c$\\
%(4260,4360,4660)$ \\
%\,\cite{BELLEY,BABARY,BES}\\
$\bar D^*_0D^*$&112&143&$157(10)$&186&238&261(17) &6020&5861&5748(101)&4328\\
$\bar DD_1$&98&126&139(13)&164&209&231(21)&5769&5639&5544(162)&4291\\
%%%%%%%%%%%%%%%%%%%%%%%%%%%%%%%%%%%%%%%%%%%%%%%%%%
$\bf { J^{PC}=1^{-+}} $ &&&&&&&&&&&$Y_c$\\
$\bar D^*_0D^*$&105&135&150(13)& 174 & 224 &249(22) &6047 & 5920 & 5828(132)&4328 \\
$\bar DD_1$&97&128&145(15)&162&213&241(25)&5973&5840&5748 (179))&-- \\
%107&140&\bf 152(10)&178&231&\bf 252(17)&6109&5954&\bf 5847(113)&4291\\
%$ J/\psi S_2$&128&175&\bf 220(19) &213&290&\bf 364(31) &5878&5643&\bf 5457(126) & 4097\\
\hline
\hline
\end{tabular*}
}}
\label{tab:d-chl}
\end{table*}
}
%%%%%%%%%%%%%%%%%%%%%%%%%%%%%%%%%%%%%%%
%%%%%%%%%%%%%%%%%%%%%%%%%%%%%%%%%%%%%%%
{\scriptsize
\begin{table*}[hbt]
\setlength{\tabcolsep}{0.8pc}
 \caption{$\bar BB$-like molecules masses, invariant and running couplings  from LSR within stability criteria from LO to N2LO of PT. The invariant coupling $\hat f_X$ is defined in Eq.\,\ref{eq:fhat}.}
% space before first and after last column: 1.5pc
% space between columns: 3.0pc (twice the above)
% -----------------------------------------------------
% adapted from TeX book, p. 241
%\newlength{\digitwidth} \settowidth{\digitwidth}{\rm 0}
%\catcode`?=\active \def?{\kern\digitwidth}
% -----------------------------------------------------
{\scriptsize{
%\begin{tabular}{\textwidth}{@{}l@{\extracolsep{\fill}}  lll   lll  lll  c}
\begin{tabular*}{\textwidth}{@{}lll   lll  lll  l cc@{\extracolsep{\fill}}l}
%\begin{tabular}{lll   lll  lll  l c}
\\
\hline
\hline
%\\
                 \bf Nature& \multicolumn{3}{c}{${\hat f_X}$  [keV]} 
                 & \multicolumn{3}{c }{${ f_X(5.5)}$  [keV]} 
                 &  \multicolumn{3}{c}{ Mass  [MeV]} 
                  & Threshold
                 &  Exp.
                 \\
\cline{2-4} \cline{5-7}\cline{8-10}
%\\
                 & \multicolumn{1}{l}{LO} 
                 & \multicolumn{1}{l}{NLO} 
                 & \multicolumn{1}{l }{ N2LO} 
                      & \multicolumn{1}{l}{LO} 
                 & \multicolumn{1}{l}{NLO} 
                 & \multicolumn{1}{l }{ N2LO} 
                   & \multicolumn{1}{l}{LO} 
                 & \multicolumn{1}{l}{NLO} 
                 & \multicolumn{1}{l}{ N2LO} 
                  \\
%\\
\hline
%\\
 $\bf{ J^{PC}=0^{++}}$ &&&&&&&&&&&--\\
$\bar BB$&4.0&4.4&5(1)&14.4&15.6&17(4)&10605&10598&10595(58)&10559\\
$\bar B^*B^*$&--&--&--&27&30&32(5)&10626&10646&10647(184)&10650\\
$ B^*_0B^*_0$&2.1&3.2&4(1)&7.7&11.3&14(4)&10653&10649&10648(113)&--\\
$\bar B_{1}B_{1}$&--&--&--&--&20&28.6(4)&--&10514&10514(149)&--\\
\\
$\bf { J^{PC}=1^{+\pm} }$ &&&&&&&&&&&$X_b,Z_b$\\
%(10607,10652)$\\
%\,\cite{BELLEZb}
$\bar B^*B$&7&8&9(3)&14&16&17(5)&10680&10673&10646(150)&10605\\
$\bar B^*_0B_1$&4&6&7(1)&8&11&14(2)&10670&10679&10692(132)&--\\
\\
%%%%%%%%%%%%%%%%%%%%%%%%%%%%%%%%%%%%%%%%%%%%%%
$\bf { J^{PC}=0^{-\pm}} $ &&&&&&&&&&&--\\
$\bar B^*_0B$&11&16&20(3)&39&55&67(10)&12930&12737&12562(260) &--
\\
$\bar B^*B_1$&--&--&-- &71&105&136(19) &12967&12794&12627(225)&11046\\
\\
$\bf  {J^{PC}=1^{--}} $ &&&&&&&&&&& $Y_b$\\
%(9898,10260,10870)$\\
%\,\cite{BELLEYb1,BELLEYb2}\\
$\bar B^*_0B^*$&21&29&35(6) &39&54&66(11) &12936&12756&12592(266) &--\\
$\bar BB_1$&21&29&35(7)&39&54&65(12)&12913&12734&12573(257)&11000\\
%%%%%%%%%%%%%%%%%%%%%%%%%
$\bf  {J^{PC}=1^{-+}} $ &&&&&&&&&&& $Y_b$\\
$\bar B^*_0B^*$&20&29&34(4)&38&54&64(8)&12942&12774&12617(220) &-- \\
$\bar BB_1$& 20&29&35(5)&	37&53&65(9) &12974& 12790&12630(236)					&11000\\
%21&29&\bf 35(5)&38&54&\bf 64(9)&12979&12797&\bf 12637(224)&11000\\
%$ \Upsilon S_2$&11&14&\bf 15(2) &21&27&\bf 27(3)&11757&11597&\bf 11562(185)& 10460\\
%\\
\hline
\hline
\end{tabular*}
}}
\label{tab:b-chl}
\end{table*}
}
%%%%%%%%%%%%%%%%%%%%%%%%%%%%%%%%%%%%%%%
%%%%%%%%%%%%%%%%%%%%%%%%%%%%%%%%%%%%%%%
{\scriptsize
\begin{table*}[hbt]
\setlength{\tabcolsep}{0.9pc}
\caption{Four-quark masses, invariant and running couplings  from LSR within stability criteria from LO to N2LO of PT. The decay constant is evaluated at $4.5$ (resp. $5.5$) GeV for the $c$ (resp. $b$) channel. The invariant coupling $\hat f_X$ is defined in Eq.\,\ref{eq:fhat}.}  
% space before first and after last column: 1.5pc
% space between columns: 3.0pc (twice the above)
% -----------------------------------------------------
% adapted from TeX book, p. 241
%\newlength{\digitwidth} \settowidth{\digitwidth}{\rm 0}
%\catcode`?=\active \def?{\kern\digitwidth}
% -----------------------------------------------------
%\label{tab:4q-result}
{\scriptsize{
%\begin{tabular}{\textwidth}{@{}l@{\extracolsep{\fill}}  lll   lll  lll  c}
\begin{tabular*}{\textwidth}{@{}lll   lll  lll  l cc@{\extracolsep{\fill}}l}
%\begin{tabular}{lll   lll  lll  l c}
\\
\hline
\hline
%\\
                \bf Nature& \multicolumn{3}{c}{{$\hat f_X$}  [keV]} 
                 & \multicolumn{3}{c }{{ $f_X$}  [keV]} 
                 &  \multicolumn{3}{c}{ Mass  [MeV]} 
                 & Exp.
                 \\
\cline{2-4} \cline{5-7}\cline{8-10}
%\\
                 & \multicolumn{1}{l}{LO} 
                 & \multicolumn{1}{l}{NLO} 
                 & \multicolumn{1}{l }{ N2LO} 
                      & \multicolumn{1}{l}{LO} 
                 & \multicolumn{1}{l}{NLO} 
                 & \multicolumn{1}{l }{ N2LO} 
                   & \multicolumn{1}{l}{LO} 
                 & \multicolumn{1}{l}{NLO} 
                 & \multicolumn{1}{l}{ N2LO} 
                  \\
%\\
\hline
 \bf{$c$}\bf-quark \\
 $S_c(0^{+})$&62&67&70(7)&173&184&191(20)&3902&3901&3898(54)&--  \\
 $A_c(1^{+})$&100&106&112(18)&166&176&184(30)&3903&3890&3888(130)&$X_c,Z_c$\\%(3872)$, $Z_c(3900,4430)$  \\
 $\pi_c(0^{-})$&84&106&113(5)&233&292&310(13)&6048&5872&5750(127)&--  \\
 $V_c(1^{-})$&123&162&178(11)&205&268&296(19)&6062&5904&5793(122)&$Y_c$\\%(4260,4360,4660)$ \\
 \\
 \bf {$b$}\bf-quark \\
 $S_b(0^{+})$&4.6&5.0&5.3(1.1)&16&17&19(4)&10652&10653&10654(109)&--  \\
 $A_b(1^{+})$&8.7&9.5&10(2)&16&18&19(3)&10730&10701&10680(172)&$Z_b$\\ %(10607,10652)$ \\
 $\pi_b(0^{-})$&18&23&27(3)&62&83&94(11)&13186&12920&12695(254)&--  \\
 $V_b(1^{-})$&24&33&40(5)&45&62&75(9)&12951&12770&12610(242)&$Y_b$\\%(9898,10260,10870)$ \\
\hline
\hline
\end{tabular*}
}}
\label{tab:4q-chl}
\end{table*}
}
%%%%%%%%%%%%%%%%%%%%%%%%%%%%%%%%%%%%%%%
%%%%%%%%%%%%%%%%%%%%%%%%%%%%%%%%%%%%%%%
%%%%%%%%%%
\subsubsection*{$\bullet$ $\bar{D}D$-SU3 breaking ($\bar{D}_sD_s$)}
\begin{figure}[hbt] 
\begin{center}
{\includegraphics[width=3.8cm  ]{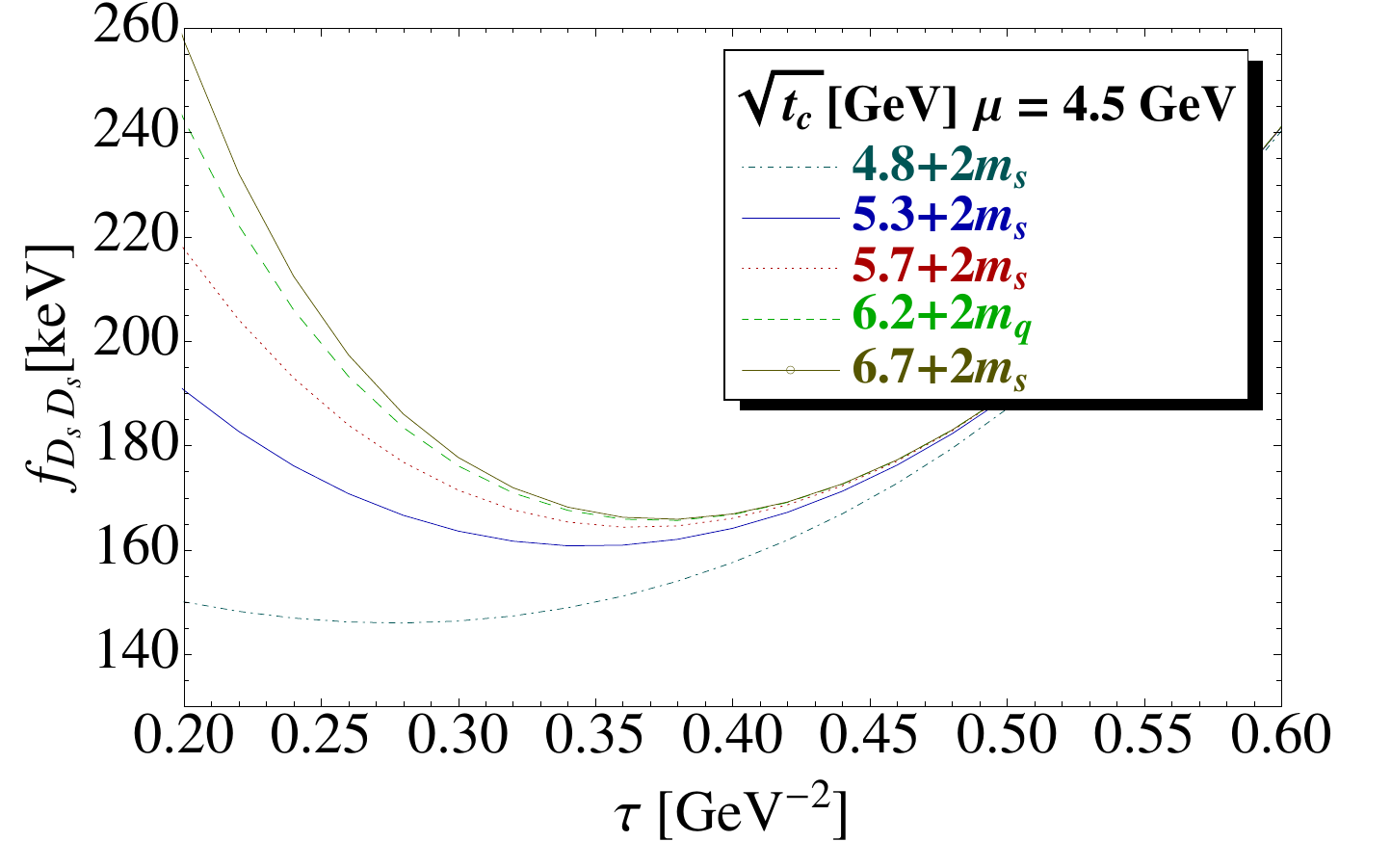}}
{\includegraphics[width=3.8cm  ]{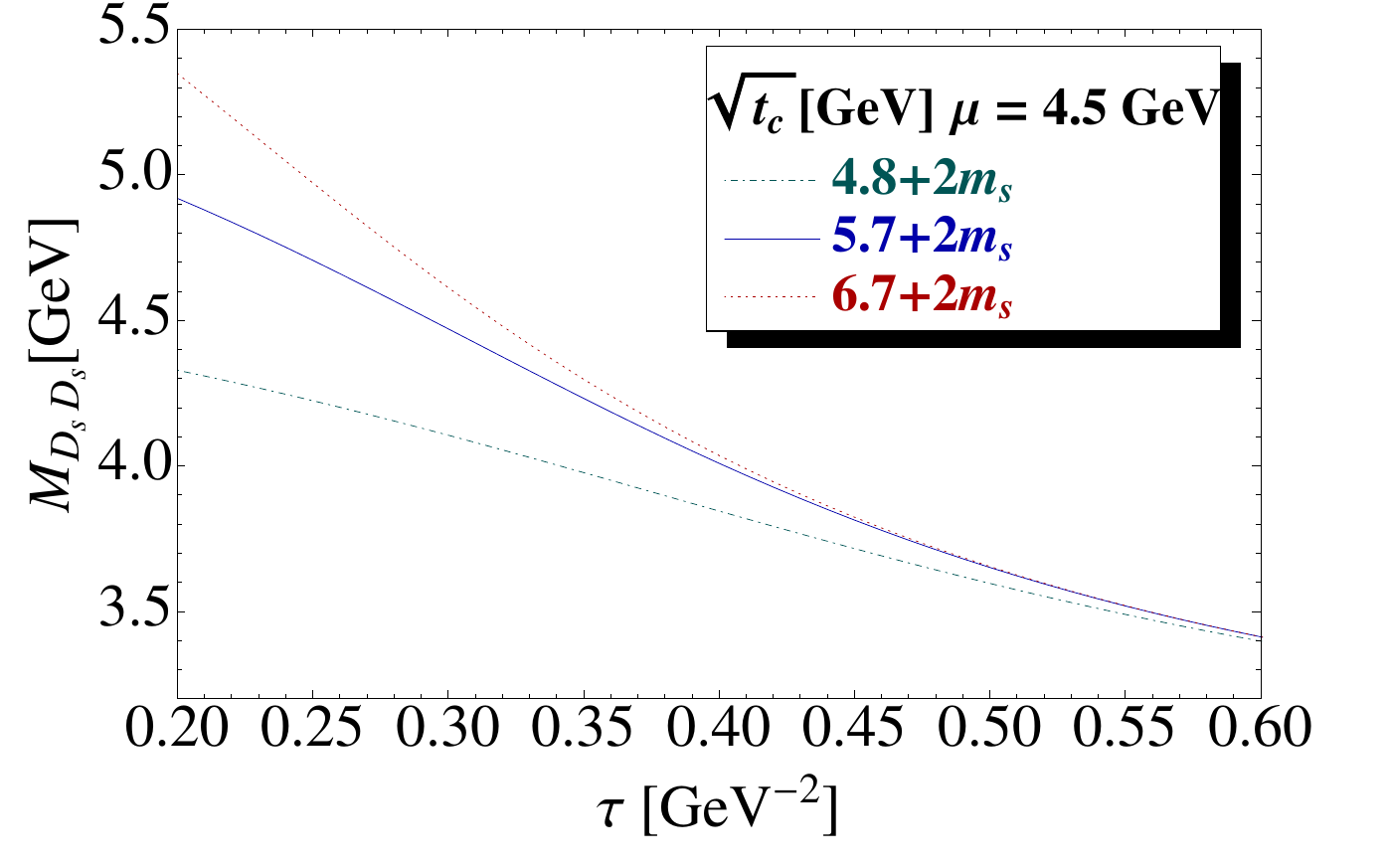}}
\centerline {\hspace*{-0.5cm} a)\hspace*{3cm} b) }
\caption{
\scriptsize 
{\bf a)} The coupling $f_{D_s D_s}$  at NLO as function of $\tau$ for different values of $t_c$, for $\mu=4.5$ GeV  and for the QCD parameters in Table\,\ref{tab:param}; {\bf b)} The same as a) but for the mass $M_{D_s D_s}$.
}
\label{fig:dsds-nlo} 
\end{center}
\end{figure}

%%%%%%%%%%
The analysis of the $\mu$ subtraction point is very similar to the previous one and will not be repeated here. Using the optimal choice $\mu=4.5~\mbox{GeV}$ obtained previously, we study the behaviour of the coupling and mass in term of LSR variable $\tau$ for different values of $t_c$ at NLO (see Fig.\,\ref{fig:dsds-nlo}).
%%%%%%%%%%%%%%%%%%%%%%%%%
\nin
%%%%%%%%%%
\subsubsection*{-- SU3 ratios of masses and couplings}
\begin{figure}[hbt] 
\begin{center}
{\includegraphics[width=3.8cm  ]{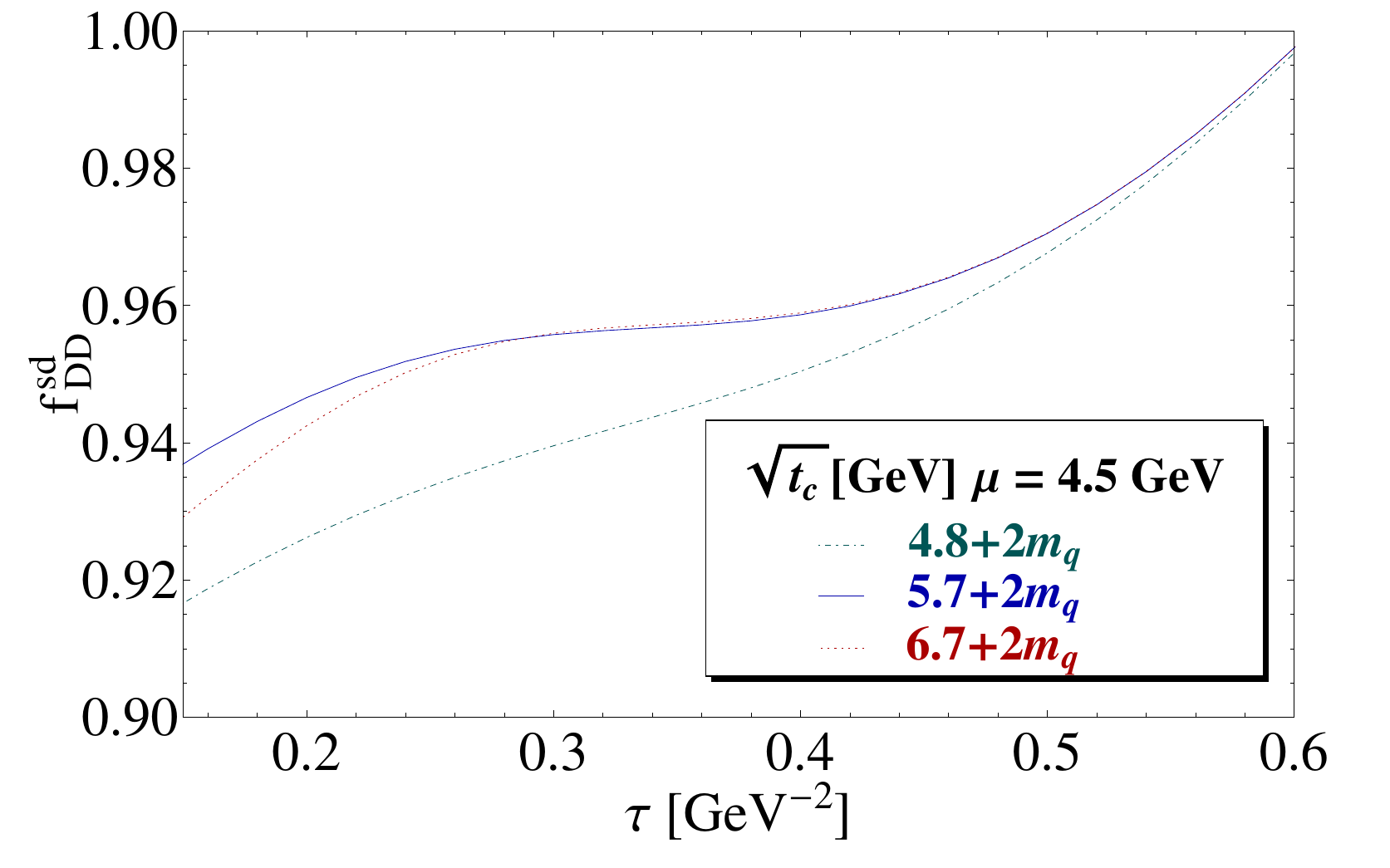}}
{\includegraphics[width=3.8cm  ]{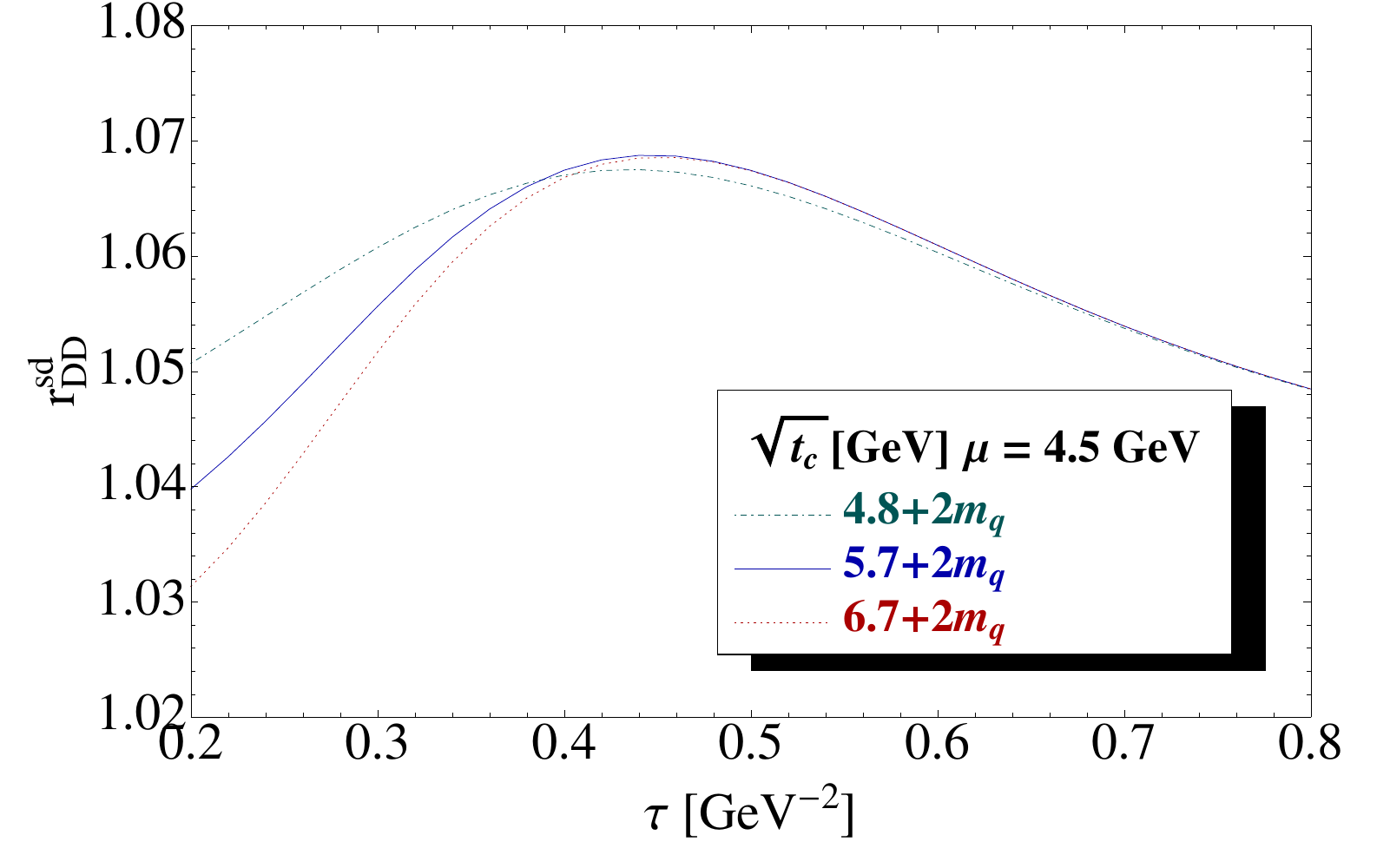}}
\centerline {\hspace*{-0.5cm} a)\hspace*{3cm} b) }
\caption{
\scriptsize 
{\bf a)} The SU3 ratio of couplings $f^{sd}_{DD}$ at NLO  as function of $\tau$ for different values of $t_c$, for $\mu=4.5$ GeV  and for the QCD parameters in Table\,\ref{tab:param}; {\bf b)} The same as a) but for the ratio of mass $r^{sd}_{DD}$.
}
\label{fig:frsd-nlo} 
\end{center}
\end{figure}

%%%%%%%%%%
We study the behaviour of the SU3 ratios of couplings ($f^{sd}_{DD}$) and masses ($r^{sd}_{DD}$) in terms of LSR variable $\tau$ for different values of $t_c$ at NLO as shown in Fig.\,\ref{fig:frsd-nlo}.
%%%%%%%%%%%%%%%%%%%%%%%%%
\nin
%%%%%%%%%%%%%%%%%%%%%%%%%%%%
We consider as an optimal estimate the mean value of the coupling, mass and their SU3 ratios obtained at the minimum or inflexion point for the common range of $t_c$-values ($\sqrt{t_c}\simeq 4.8+2\overline{m}_s ~\mbox{GeV}$ for $\tau\simeq (0.28\pm 0.02)~ \mbox{GeV}^{-2}$) corresponding to the starting of the $\tau$-stability and the one where (almost) $t_c$-stability ($\sqrt{t_c}\simeq 6.7+2\overline{m}_s ~\mbox{GeV}$) is reached for $\tau\simeq (0.38\pm 0.02)~ \mbox{GeV}^{-2}$. %(Fig. \ref{fig:dsds-nlo} a))
%%%%%%%%%%%%%%%%%%%%%%%%%%%%%%%%%%%%%%%
%%%%%%%%%%%%%%%%%%%%%%%%%%%%%%%%%%%%%%%
%%%   RESULTS SU3 BREAKING         %%%%
%%%%%%%%%%%%%%%%%%%%%%%%%%%%%%%%%%%%%%%
%%%%%%%%%%%%%%%%%%%%%%%%%%%%%%%%%%%%%%%
%%%%%%%%%%%%%%%%%%%%%%%%%%%%%%%%%%%%%%%
%%%%%%%%%%%%%%%%%%%%%%%%%%%%%%%%%%%%%%%
{\scriptsize
\begin{table*}[hbt]
\setlength{\tabcolsep}{1.13pc}
 \caption{$\bar{D}D$-like molecules couplings, masses and their corresponding SU3 ratios  from LSR within stability criteria at NLO to N2LO of PT.}% We include revised estimates of the $ \bar D^{*}_{0}D^{*}_{0}$,  $ \bar D^{*}_{0}D_{1}$ couplings and masses and new one for $ \bar D_{1}D_{1}$.}  
{\scriptsize
\begin{tabular*}{\textwidth}{@{}ll   ll  ll  ll l@{\extracolsep{\fill}}l}
%\begin{tabular*}{\textwidth}{@{}l@{\extracolsep{\fill}}l}
\\
\hline
\hline
%\\
                \bf Channels &\multicolumn{2}{c}{$f^{sd}_M\equiv f_{M_s}/f_{M}$}
					&\multicolumn{2}{c}{$f_{M_s}(4.5)$ [keV]}
					&\multicolumn{2}{c}{$r^{sd}_M\equiv M_{M_s}/M_{M}$}
					&\multicolumn{2}{c}{$M_{M_s}$  [MeV]}\\
\cline{2-3} \cline{4-5}\cline{6-7}\cline{8-9}
%\\
                 & \multicolumn{1}{l}{{NLO}}
                 & \multicolumn{1}{l }{N2LO} 
                 & \multicolumn{1}{l}{NLO} 
                 & \multicolumn{1}{l }{N2LO} 
                 & \multicolumn{1}{l}{NLO} 
                 & \multicolumn{1}{l}{N2LO}
		    & \multicolumn{1}{l}{NLO} 
                 & \multicolumn{1}{l}{N2LO} 
                  \\
%\\
\hline
%\\
 \bf{Scalar($0^{++}$)}&&&&&&&&\\
$\bar D_sD_s$&$0.95(3)$&0.98(4)&156(17)&167(18)&1.069(4)&1.070(4)&4169(48)&4169(48)\\
$\bar D^{*}_{s}D^{*}_{s}$&0.93(3)&0.95(3)&265(31)&284(34)&1.069(3)&1.075(3)&4192(200)&4196(200)\\
$\bar D^{*}_{s0}D^{*}_{s0}$&0.88(6)&0.89(6)&85(12)&102(14)&1.069(69)&1.058(68)&4277(134)&4225(132)\\
$\bar D_{s1}D_{s1}$&0.906(33)&0.930(34)&209(28)&229(31)&1.097(7)&1.090(7)&4187(62)&4124(61) \\
\\
%$\bar D^{*}_{0}D^{*}_{0}$&--&--&97(15)&114(18)&--&--&4003(227)&3954(224)\\
%$\bar D_{1}D_{1}$&--&--&236(32)&274(37)&--&--&3838(57)&3784(56) \\
\bf {Axial($1^{+\pm}$)}&&&&&&&&\\
$\bar D^{*}_{s}D_{s}$&0.93(3)&0.97(3)&143(16)&156(17)&1.070(4)&1.073(4)&4174(67)&4188(67)\\
$\bar D^{*}_{s0}D_{s1}$&0.90(1)&0.82(1)&87(14)&110(18)&1.119(24)&1.100(24)&4269(205)&4275(206)\\
%\\
%$\bar D^{*}_{0}D_{1}$&--&--&96(15)&112(17)&--&--&3849(182)&3854(182)\\
%%%%%%%%%%%%%%%%%%%%%%%%%%%%%%%%%%%%%%%%%%%%%%%%%%%
\bf  {Pseudo($0^{-\pm}$)}&&&&&&&&\\
$\bar D^{*}_{s0}D_{s}$&0.94(5)&0.90(4)&225(24)&232(25)&0.970(50)&0.946(40)&5604(223)&5385(214)\\
$\bar D^{*}_{s}D_{s1}$&0.93(4)&0.90(4)&455(34)&508(38)&0.970(50)&0.972(34)&5724(195)&5632(192)\\
%%%%%%%%%%%%%%%%%%%%%%%%%%%%%%%%%%%%%%%%%%%%%%%%%%
%\\
\bf {Vector($1^{--}$)} &&&&&&&&\\
$\bar D^{*}_{s0}D^{*}_{s}$&0.87(4)&0.86(4)&208(11)&216(11)&0.980(33)&0.956(32)&5708(184)&5571(180)\\
$\bar D_{s}D_{s1}$&0.97(3)&0.93(3)&202(12)&213(13)&0.970(33)&0.951(31)&5459(122)&5272(120)\\
%%%%%%%%%%%%%%%%%%%%%%%%%%%%%%%%%%%%%%%%%%%%%%%%%%
\bf { Vector($1^{-+}$)} &&&&&&&&\\
$\bar D^{*}_{s0}D^{*}_{s}$&0.98(5)&0.92(5)&219(17)&231(18)&0.963(32)&0.948(32)&5699(184)&5528(179)\\
$\bar D_{s}D_{s1}$&0.92(3)&0.88(3)&195(13)&212(14)&0.959(34)&0.955(34)&5599(155)&5487(152)\\
%\\
\hline
\hline
%\\
%\hline
%\hline
\end{tabular*}
}
\label{tab:d-su3}
\end{table*}
}
%%%%%%%%%%%%%%%%%%%%%%%%%%%%%%%%%%%%%%%
%%%%%%%%%%%%%%%%%%%%%%%%%%%%%%%%%%%%%%%
{\scriptsize
\begin{table*}[h]
\setlength{\tabcolsep}{1.08pc}
 \caption{$\bar{B}B$-like molecules couplings, masses and their corresponding SU3 ratios  from LSR within stability criteria at NLO to N2LO of PT.  The * indicates that the value does not come from a direct determination.
}
{\scriptsize
\begin{tabular*}{\textwidth}{@{}ll   ll  ll  ll l@{\extracolsep{\fill}}l}
%\begin{tabular}{@{}ll   ll  ll  ll l@{}}
\\
\hline
\hline
%\\
                \bf Channels &\multicolumn{2}{c}{$f^{sd}_M\equiv f_{M_s}/f_{M}$}
					&\multicolumn{2}{c}{$f_{M_s}(5.5)$ [keV]}
					&\multicolumn{2}{c}{$r^{sd}_M\equiv M_{M_s}/M_{M}$}
					&\multicolumn{2}{c}{$M_{M_s}$  [MeV]}\\
\cline{2-3} \cline{4-5}\cline{6-7}\cline{8-9}
%\\
                 & \multicolumn{1}{l}{NLO} 
                 & \multicolumn{1}{l }{N2LO} 
                 & \multicolumn{1}{l}{NLO} 
                 & \multicolumn{1}{l }{N2LO} 
                 & \multicolumn{1}{l}{NLO} 
                 & \multicolumn{1}{l}{N2LO}
		    & \multicolumn{1}{l}{NLO} 
                 & \multicolumn{1}{l}{N2LO} 
                  \\
%\\
\hline
%\\
\bf{Scalar($0^{++}$)}&&&&&&&&\\
$\bar B_sB_s$&1.04(4)&1.15(4)&17(2)&20(2)&1.027(4)&1.029(4)&10884(74)&10906(74)\\
$\bar B^{*}_{s}B^{*}_{s}$&1.00(3)&1.12(3)&31(5)&36(6)&1.028(5)&1.029(5)&10944(134)&10956(134)\\
$\bar B^{*}_{s0}B^{*}_{s0}$&1.11(5)&1.07(5)&13(3)&17(4)&1.050(11)&1.034(11)&11182(227)&11014(224)\\
$\bar B_{s1}B_{s1}$&1.197(73)&1.214(74)&24(5)&29(6)&1.040(2)&1.035(2)&10935(170)&10882(169) \\
 \\
%$\bar B_{1}B_{1}$&--&--&20(3)&28.6(4)&--&--&10514(149)&10514(149) \\
\bf {Axial($1^{+\pm}$)}&&&&&&&&\\
$\bar B^{*}_{s}B_{s}$&1.01(3)&1.18(4)&16.7(2)&20(2)&1.028(4)&1.030(4)&10972(195)&10972(195)\\
$\bar B^{*}_{s0}B_{s1}$&0.80(4)&0.79(4)&9.1(2.2)&10.7(2.6)&1.052(14)&1.031(14)&11234(208)&11021(204)\\
%%%%%%%%%%%%%%%%%%%%%%%%%%%%%%%%%%%%%%%%%%%%%%%%%%%
%\\
\bf  {Pseudo($0^{-\pm}$)}&&&&&&&&\\
$\bar B^{*}_{s0}B_{s}$&1.06(3)&1.02(3)&58(3)&68(4)&1.00(3)*&1.00(3)*&12725(217)&12509(213)\\
$\bar B^{*}_{s}B_{s1}$&0.96(4)&0.95(4)&100(11)&118(13)&1.00(3)*&1.00(3)*&12726(295)&12573(292)\\
%%%%%%%%%%%%%%%%%%%%%%%%%%%%%%%%%%%%%%%%%%%%%%%%%%
%\\
\bf { Vector($1^{--}$)} &&&&&&&&\\
$\bar B^{*}_{s0}B^{*}_{s}$&0.95(3)&0.90(3)&51(4)&59(5)&1.00(3)*&0.99(3)*&12715(267)&12512(263)\\
$\bar B_{s}B_{s1}$&0.83(4)&0.77(3)&45(3)&50(3)&0.99(3)*&0.99(3)*&12615(236)&12426(233)\\
%%%%%%%%%%%%%%%%%%%%%%%%%%%%%%%%%%%%%%%%%%%%%%%%%%
\bf { Vector($1^{-+}$)} &&&&&&&&\\
$\bar B^{*}_{s0}B^{*}_{s}$&0.94(3)&0.92(3)&51(5)&59(6)&1.00(3)*&0.99(3)*&12734(262)&12479(257)\\
$\bar B_{s}B_{s1}$&0.89(4)&0.85(3)&48(5)&55(6)&0.99(3)*&0.98(3)*&12602(247)&12350(242)\\
%\\
\hline
\hline
\end{tabular*}
}
\label{tab:b-su3}
\end{table*}
}
%%%%%%%%%%%%%%%%%%%%%%%%%%%%%%%%%%%%%%%
%%%%%%%%%%%%%%%%%%%%%%%%%%%%%%%%%%%%%%%
{\scriptsize
\begin{table*}[hbt]
\setlength{\tabcolsep}{1.13pc}
 \caption{Four-quark couplings, masses and their corresponding SU3 ratios  from LSR within stability criteria at NLO and N2LO of PT. The * indicates that the value does not come from a direct determination. The decay constant is evaluated at $4.5$ (resp. $5.5$) GeV for the $c$ (resp. $b$) channel.}  
{\scriptsize{
\begin{tabular*}{\textwidth}{@{}ll   ll  ll  ll l@{\extracolsep{\fill}}l}
%\begin{tabular}{@{}ll   ll  ll  ll l@{}}
\\
\hline
\hline
%\\
                \bf Channels &\multicolumn{2}{c}{$f^{sd}_M\equiv f_{M_s}/f_{M}$}
					&\multicolumn{2}{c}{$f_{M_s}$ [keV]}
					&\multicolumn{2}{c}{$r^{sd}_M\equiv M_{M_s}/M_{M}$}
					&\multicolumn{2}{c}{$M_{M_s}$  [MeV]}\\
\cline{2-3} \cline{4-5}\cline{6-7}\cline{8-9}
%\\
                 & \multicolumn{1}{l}{{NLO}}
                 & \multicolumn{1}{l }{N2LO} 
                 & \multicolumn{1}{l}{NLO} 
                 & \multicolumn{1}{l }{N2LO} 
                 & \multicolumn{1}{l}{NLO} 
                 & \multicolumn{1}{l}{N2LO}
		    & \multicolumn{1}{l}{NLO} 
                 & \multicolumn{1}{l}{N2LO} 
                  \\
%\\
\hline
%\\
 \bf{c-quark}&&&&&&&&\\
$S_{sc}(0^+)$&0.91(4)&0.98(4)&161(17)&187(19)&1.085(11)&1.086(11)&4233(61)&4233(61)\\
$A_{sc}(1^+)$&0.80(4)&0.87(4)&141(15)&160(17)&1.081(4)&1.082(4)&4205(112)&4209(112)\\
$\pi_{sc}(0^-)$&0.88(7)&0.86(7)&256(29)&267(30)&0.97(3)*&0.96(3)*&5671(181)&5524(176)\\
$V_{sc}(1^-)$&0.91(10)&0.87(10)&245(31)&258(33)&0.96(4)*&0.96(4)*&5654(239)&5539(234)\\
\\
\bf{b-quark}&&&&&&&&\\
$S_{sb}(0^+)$&0.78(3)&0.83(3)&22(5)&26(6)&1.044(4)&1.048(4)&11122(149)&11133((149)\\
$A_{sb}(1^+)$&0.92(3)&0.98(3)&22(4)&26(5)&1.042(6)&1.046(6)&11150(172)&11172(172)\\
$\pi_{sb}(0^-)$&0.80(7)&0.76(4)&66(12)&71(13)&0.985(2)*&0.975(2)*&12730(215)&12374(209)\\
$V_{sb}(1^-)$&0.97(6)&0.90(6)&64(8)&68(9)&0.996(3)*&0.984(30)*&12716(272)&12411(266)\\
%\\
\hline
\hline
\end{tabular*}
}}
\label{tab:4q-su3}
\end{table*}
}
%%%%%%%%%%%%%%%%%%%%%%%%%%%%%%%%%%%%%%%
%%%%%%%%%%%%%%%%%%%%%%%%%%%%%%%%%%%%%%%
%%%   RESULTS X(5568)-LIKE         %%%%
%%%%%%%%%%%%%%%%%%%%%%%%%%%%%%%%%%%%%%%
%%%%%%%%%%%%%%%%%%%%%%%%%%%%%%%%%%%%%%%
%%%%%%%%%%%%%%%%%%%%%%%%%%%%%%%%%%%%%%%
%%%%%%%%%%%%%%%%%%%%%%%%%%%%%%%%%%%%%%%
{\scriptsize
\begin{table*}[hbt]
 \caption{\footnotesize $\bar BK-$ and $(\overline{cu})(ds)$)-like masses and couplings from LSR at N2LO. The running coupling $f_X$ is evaluated at 2-2.5 (resp. 4.5) GeV for the $c$ (resp. $b$) channel. The invariant coupling $\hat f_X$ is defined in Eq.\,\ref{eq:fhat}.}
%\tbl{
%}
\setlength{\tabcolsep}{3pc}
    {\scriptsize
\begin{tabular*}{\textwidth}{@{}lllll@{\extracolsep{\fill}}l}
%{\begin{tabular}{@{}lllll@{}}% \toprule
\\
\hline
\hline
%\\
\bf Nature&\bf$J^{P}$& Mass [MeV] &  $\hat f_X$ [keV] &  $ f_X(4.5)$ [keV]  \\
%\\
\hline
%\\
{\bf  $b$-quark channel}&\\
{\it Molecule} &&&\\
$\bar B^*K$&$1^{+}$&$5186\pm 13$ &$4.48\pm 0.92$&$8.02\pm 1.64$ \\
$\bar BK$&$0^{+}$&$5195\pm 15$&$2.58\pm 0.55$&$8.26\pm 1.76$\\
$\bar B^*_s\pi$&$1^{+}$&$5198\pm 17$&$5.32\pm 0.87$&$9.51\pm 1.55$\\
$\bar B_s\pi$&$0^{+}$&$5202\pm 24$&$3.04\pm 0.54 $&$9.74\pm 1.74$\\
{\it Four-quark $(bu)(\overline{ds})$} &&&\\
$A_b$&$1^{+}$&$5186\pm 16$&$5.05\pm 1.03$&$9.04\pm 1.84$\\
$S_b$&$0^{+}$&$5196\pm 17$&$2.98\pm 0.57$&$9.99\pm 1.90$\\
\\
{\bf  $c$-quark channel}&\\
{\it Molecule}\\
$\bar D^*K$&$1^{+}$&$2395\pm 44$ &$155\pm 29$&$226\pm 42$ \\
$\bar DK$&$0^{+}$&$2402\pm 42$&$139\pm 24$&$254\pm 48$\\
$\bar D^*_s\pi$&$1^{+}$&$2401\pm 48$ &$206\pm 31$&$277\pm 41$ \\
$\bar D_s\pi$&$0^{+}$&$ \pm 36$&$ \pm 24$&$ \pm 46$\\
%\\
{\it Four-quark $(cu)(\overline{ds})$} &&&\\
$A_c$&$1^{+}$&$2400\pm 46$&$192\pm 37$&$260\pm 50$\\
$S_c$&$0^{+}$&$2395\pm 67$&$122\pm 21$ &$221\pm 39$\\
\hline
\hline
\end{tabular*}
\label{tab:xres}
}
\end{table*}
} 
%%%%%%%%%%%%%%%%%%%%%%%%%%%%%%%%%%%%%%%%%%%
%%%%%%%%%%%%%%%%%%%%%%%%%%%%%%%%%%%%%%%%%%%
%%%%%%%%%%%%%%%%%%%%%%%%%%%%%%%%%%%%%%%%%%%
\section{Confrontation with some LO results and data}
%%%%%%%%%%%%%%%%%%%%%%%%%%%%%%%%%%%%%%%%%%%
%%%%%%%%%%
\subsection*{$\bullet$ Comparison with some previous LO QSSR results}
%%%%%%%%%%
\nin
The comparison is only informative as it is known that the LO results suffer from the ill-defined definition of the heavy quark mass used in the analysis at this order. Most of the authors (see e.g \, \cite{CHINESE2,RABN,WLZ,QT,CHINESE3}) use the running mass value which is not justified when one implicitly uses the QCD expression obtained within the on-shell scheme. The difference between some results is also due to the way for extracting the optimal information from the analysis. Here we use well-defined stability criteria verified from the example of the harmonic oscillator in quantum mechanics and from different well-known hadronic channels \cite{SNB1,SNB2}.
%%%%%%%%%%
\subsection*{$\bullet$ Confrontation with experiments}
%%%%%%%%%%
We conclude from the previous analysis that:\\
-- For the chiral limit case, one can notice that the masses of the $J^P = 0^+, 1^+$ states are most of them below the corresponding $DD$, $BB$-like thresholds and are compatible with some of the observed XZ masses, suggesting that these states can be interpreted as almost pure molecules or/and four-quark states.\\
-- The masses predictions of $\bar{D}D$, and $\bar{D}^*D^*$ molecules are compatible with the $0^{++}$ $Z_c(3900)$ experimental candidate .\\
%This theoretical predictions are far above the corresponding hadronic threshold which suggest that they might not be bound states and are difficult to separate from backgrounds, results inline with the ones of \cite{•}
-- The interpretation of the $0^{++}$ candidates as pure four-quark ground states is not favoured by our result.\\
-- The $0^{++}$ X(4700) experimental candidate might be identified with a $\bar D^*_{s0}D^*_{s0}$ molecule ground state.\\
-- The $1^{++}$ X(4147) and X(4273) are compatible within the error with the one of the $\bar D^*_sD_s$ molecule state and with the one of the axial-vector $A_c$ four-quark state.\\
-- Our predictions suggest the presence of $0^{++}$ $\bar D_sD_s$ and $\bar D^*_sD^*_s$ molecule states in the range $(4121\sim 4396)$ MeV and a $\bar D^*_{s0}D_{s1}$ state around 4841 MeV.\\
-- The predictions for the $J^P=0^-,~1^-$  non-strange states are about $1.5~\mbox{GeV}$ higher than the observed $Y_c$ mesons masses and ($1.7\sim 2.6$) $\mbox{GeV}$ higher than the observed $Y_b$ ones. Our reults do not favor their interpretation as pure molecule or/and four-quark states.\\
-- We also present new predictions for the $0^{-\pm}$, $1^{-\pm}$ and for different beauty states which can be tested in future experiments.
%\vspace*{-0.1cm} 
%%%%%%%%%%%%%%%%%%%%%%%%%%%%%%%%%%%%%%%%%%%
\section{$\bar BK$ and $(\overline{Qq})(us)$-like states and the $X(5568)$}
%%%%%%%%%%%%%%%%%%%%%%%%%%%%%%%%%%%%%%%%%%%
%\vspace*{-0.05cm}
In \cite{SNX1}, we have also studied the $X$ hadron formed by 3 light quarks $uds$ and one heavy quark $Q \equiv c, b$ using the same approach as above by assuming that it is a molecule or four-quark state. We have included NLO
and N2LO PT corrections and the contributions of condensates of dimension $d\leq 7$. Our results are summarized in Table \ref{tab:xres}. Contrary to previous claims in the sum rule literature, our results do not favour a $BK$, $B^*K$ or $B_s\pi$ molecule or four-quark $(bu)(\overline{ds})$ state having a mass around $5568$ MeV observed by D0 \cite{D0} but not confirmed by LHCb \cite{Xlhcb}. We also predict the corresponding state in the c-quark channel where the $D^*_{s0}$(2317) seen by BABAR \cite{Xbbar} in the $D_s\pi$ invariant mass, expected to be an isoscalar-scalar state with a width less than 3.8 MeV \cite{PDG} could be a good candidate for one of such states. From our analysis, one may suggest an experimental scan of the regions ($2327 \sim 2444$) $\mbox{MeV}$ and ($5173 \sim 5226$) $\mbox{MeV}$ for detecting these unmixed $(\overline{cu})ds$ and $(\overline{bu})ds$ exotic hadrons (if any) via eventually their radiative or $\pi +$ hadrons decays.
%%%%%%%%%%%%%%%%%%%%%%%%%%%%%%%%%%%%%%%%%%%
\section{Conclusion}
%%%%%%%%%%%%%%%%%%%%%%%%%%%%%%%%%%%%%%%%%%%
We have presented improved predictions of QSSR for the masses and couplings of molecules and four-quarks states at N2LO  of PT series and including non-perturbative contributions of condensates up to dimension $d\leqslant 6$. We can see a good convergence of the PT series after including higher order corrections which confirms the veracity of our results. For the charm (resp. bottom) channels, our predictions for molecules and four-quark vector non-strange states are too high compared with the observed $Y_c(4140)$ to $Y_c(4660)$ (resp. $Y_b(9898,~10260,~10870)$). We can also notice that the SU3 breakings are tiny for the masses ($\leqslant 10$ (resp. $3$)$\%$) for the charm (resp. bottom) channels but can be large for the couplings ($\leqslant 20 \%$). This can be understood as in the ratios of sum rules, the corrections tend to cancel out. Our analysis has been done within stability criteria with respect to the LSR variable $\tau$, the QCD continuum threshold $\sqrt{t_c}$ and the subtraction constant $\mu$ which have provided successful predictions in different hadronic channels \cite{SNB1,SNB2}. The optimal values of the masses and couplings have been extracted at the same value of these parameters where the stability appears as an extremum and/or inflection points. We do not include higher dimension condensates contributions (which are not under good control) in our estimate but only consider them as a source of errors. This work is an improvement of all existing previous LO results in the literature from QCD spectral sum rules on XYZ like masses and couplings. We expect that future experimental data and/or lattice results will check our predictions.
%%%%%%%%%%%%%%%%%%%%%%%%%%%%
%We have summarized  our results for SU3 breaking at NLO and N2LO of PT\,\cite{SU3}
%for  molecule and four-quark states (see Table\,\ref{tab:resultc} to Table\,\ref{tab:4q-resultb}). They are important for further building of an effective theory for these exotic states and can be tested by lattice calculations. We plan to extend this analysis for the estimate of the meson widths.
%%%%%%%%%%%%%%%%%%%%%%%%%%%%
%\section*{Acknowledgements}
%%%%%%%%%%%%%%%%%%%%%%%%%%%%
%This work is done in Collaboration with: R. Albuquerque, S. Narison and G. Randriamanatrika.
%We thank A. Rabemananjara for participating at the early stage of this work.
%%%%%%%%%%%%%%%%%%%%%%%%%%%%
%%%%%%%%%%%%%%%%%%%%%%%%%%%%
%%%%%%%%%%%%%%%%%%%%%%%%%%%%
%%%%%%%%%%%%%%%%%%%%%%%%%%%%
%\clearpage

\end{document}